%% file: main.tex
\documentclass[sigconf]{acmart}

\usepackage{needspace}
\usepackage{soul}
\AtBeginDocument{%
  }

\copyrightyear{2025}
\acmYear{2025}
\setcopyright{cc}
\setcctype[4.0]{by-sa}
\acmConference[SCF '25]{ACM Symposium on Computational Fabrication}{November 20--21, 2025}{Cambridge, MA, USA}
\acmBooktitle{ACM Symposium on Computational Fabrication (SCF '25), November 20--21, 2025, Cambridge, MA, USA}
\acmDOI{10.1145/3745778.3766654}
\acmISBN{979-8-4007-2034-5/2025/11}

\begin{document}

\title{Y-AR: A Mixed Reality CAD Tool for 3D Wire Bending}

\author{Shuo Feng}
\orcid{0000-0003-2350-321X}
\affiliation{%
  \institution{Cornell Tech}
 \city{New York}
\country{USA}
}
\email{sf522@cornell.edu}

\author{Bo Liu}
\affiliation{%
  \institution{Cornell Tech}
 \city{New York}
\country{USA}
  }
  \email{bl685@cornell.edu}

\author{Yifan (Lavenda) Shan}
\affiliation{%
  \institution{Cornell Tech}
 \city{New York}
\country{USA}
}
\email{ys2253@cornell.edu}

\author{Roy Zunder}
\affiliation{%
  \institution{Cornell Tech}
\city{New York}
\country{USA}
}
\email{rz482@cornell.edu}

\author{Wei-Che Lin}
\affiliation{%
  \institution{Cornell University}
 \city{Ithaca}
\country{USA}
}
\email{wl725@cornell.edu}

\author{Tri Dinh}
\affiliation{%
  \institution{Macaulay Honors College}
\city{New York}
\country{USA}
}
\email{tri.dinh@macaulay.cuny.edu}

\author{Harald Haraldsson}
\affiliation{%
  \institution{Cornell Tech}
\city{New York}
\country{USA}
}
\email{hh586@cornell.edu}

\author{Ofer Berman}
\authornote{Ofer was partially affiliated with Technion and Cornell Tech during this study.}
\affiliation{%
  \institution{Technion and Cornell Tech*}  
\country{Israel / New York, USA}
}
\email{ofer.berman@technion.ac.il}

\author{Thijs Roumen}
\affiliation{%
  \institution{Cornell Tech}
\city{New York}
\country{USA}
  }
\email{thijs.roumen@cornell.edu}

\renewcommand{\shortauthors}{Feng et al.}

\begin{abstract}

Wire bending is a technique used in manufacturing to mass-produce items such as clips, mounts, and braces. Recent advances in programmable wire bending have made this process increasingly accessible for custom fabrication. However, CNC wire benders are controlled using Computer Aided Manufacturing (CAM) software, without design tools, making custom designs challenging to produce. We present \textit{Y-AR}, a Computer Aided Design (CAD) interface for 3D wire bending. \textit{Y-AR} uses mixed reality to let designers create clips, mounts, and braces to physically connect objects to their surrounding environment (such as the mount shown in Figure~\ref{fig:teaser}c). The interface incorporates springs as design primitives which (1)~apply forces to hold objects, and (2)~counter-act dimensional inaccuracies inherently caused by mid-air modeling and measurement errors in AR. Springs are a natural design element when working with metal wire-bending given its specific material properties. We demonstrate workflows to design and fabricate a range of mechanisms in \textit{Y-AR} as well as structures made using free-hand design tools. We found that combining gesture-based interaction with fabrication-aware design principles allowed novice users to create functional wire connectors, even when using imprecise XR-based input. In our usability evaluation, all 12 participants successfully designed and fabricated a functional bottle holder using \textit{Y-AR}.


\end{abstract}

\begin{teaserfigure}
  \includegraphics[width=\textwidth]{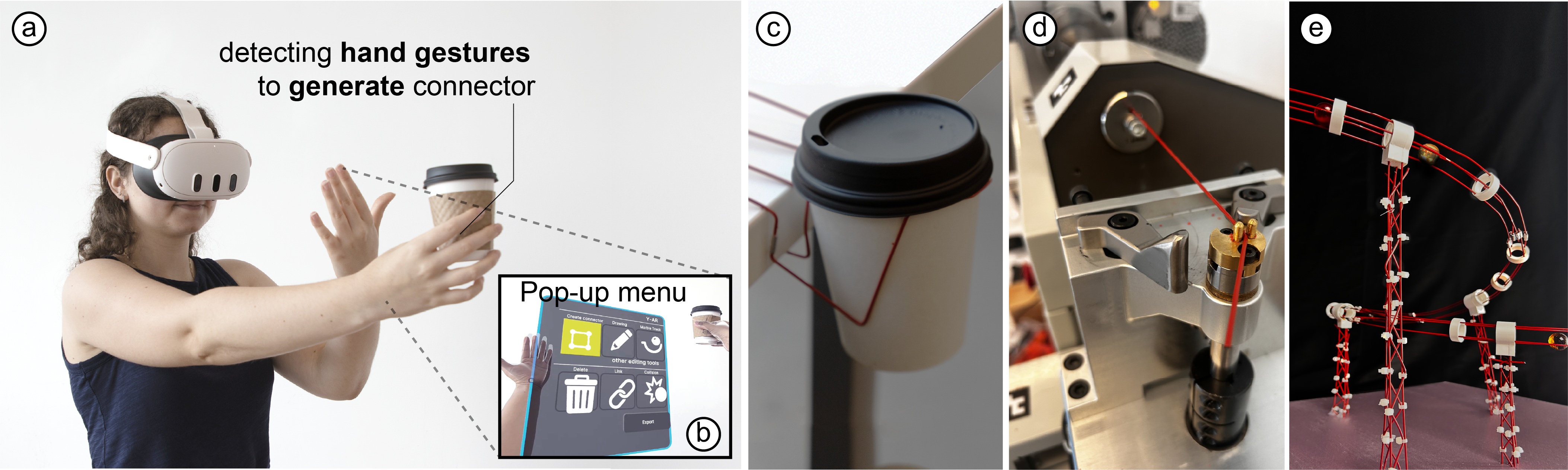}
  \caption{\textit{Y-AR} is a Computer-Aided Design~(CAD) tool for 3D wire bending. (a)~the XR interface lets users create structures that fit to real-world dimensions, (b)~it leverages hand gestures to generate connectors, such as (c)~a cup holder designed to attach to the edge of a table, or freeform structures such as a (e)~marble track. (d)~\textit{Y-AR} exports instructions for CNC wire bending machines.}
  \Description{}
  \label{fig:teaser}
\end{teaserfigure}


\makeatletter
\def\@ACM@copyright@check@cc{%
  \if@ACM@nonacm
     \ClassInfo{\@classname}{Using CC license with a non-acm
       material}%
  \else
     \if@ACM@engage
        \ClassInfo{\@classname}{Using CC license with ACM Enage
          material}%
      \else
      \fi
  \fi}
\makeatother
\maketitle

\input{src/00_INTRODUCTION.tex}

\input{src/01_RELATEDWORK.tex}

\input{src/02_USERINTERFACE.tex}

\input{src/03_DESIGNSPACE.tex}

\input{src/05_USERSTUDY.tex}

\input{src/06_DISCUSSION.tex}



\begin{acks}
We thank the XR Collaboratory at Cornell Tech for their generous sponsorship of this research in the form of the inaugural XR Prototyping Grant.  
We also acknowledge support from the BURE program at Cornell University.
We further thank Pensa Labs for generously providing access to their wire bender, which was essential for prototyping and testing the wire structures presented in this work.
\end{acks}

\bibliographystyle{ACM-Reference-Format}
\bibliography{ Reference,thijs-papers}

\input{src/08_appendix}

\end{document}

%% file: src/00_INTRODUCTION.tex
\section{Introduction}

 \begin{figure*}[h]
  \centering
  \includegraphics[width=\linewidth]{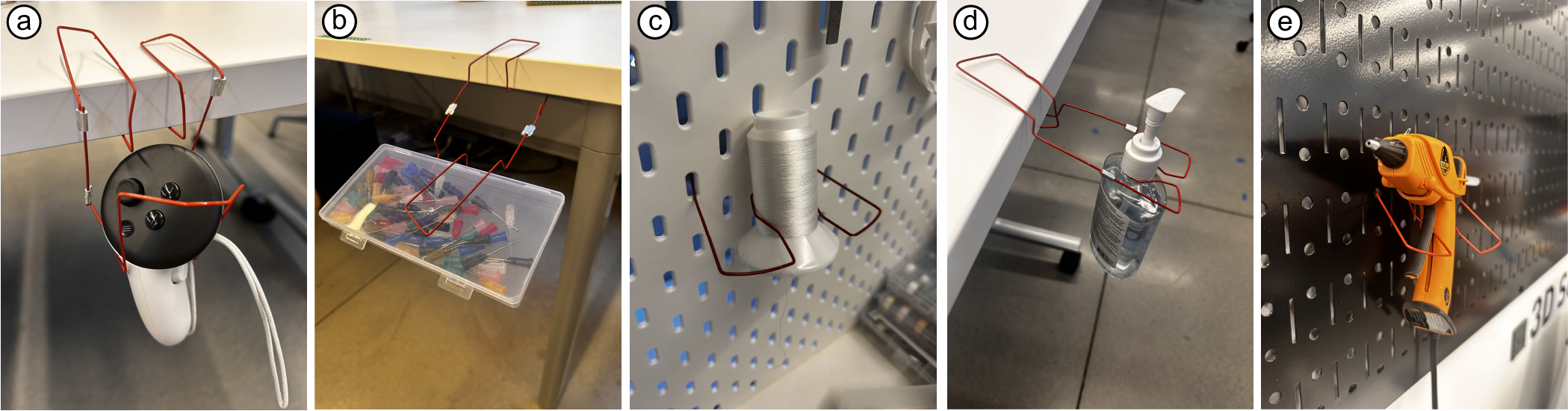}
 \caption{We designed a series of connectors, including for various objects.}
  \label{fig:demo object}
\end{figure*}

Wire bending is a common industrial mass-manufacturing process used to create springs, clips, mounts, and braces, among others. 
In sculpting, wire bending is used to provide internal support for incomplete structures, for example, as an \textit{armature}. While most wire bending applications are in \textit{high-volume, low-mix} mass production, nascent CNC wire bending machines enable \textit{on-demand} wire bending of \textit{high-mix, low volume} objects. This mode of production is essential for wire-bending in the medical sector, where it is used to create retainers and braces~\cite{alassiry2019orthodontic}, as well as in creative domains such as puppetry, where it is used to produce actuated characters~\cite{BENDIT}. To date, most of these applications rely on manual workflows where wires are bent by hand.

New machines such as the DIWire by Pensalabs make automated wire bending accessible for personal fabrication~\cite{Baudisch_Mueller_2017}. Hobbyists have embraced it to create connectors, holders, signs, marble tracks, and sculptures. While software exists to operate these machines (e.g., \textit{VGPNEXT}\footnote{https://www.blmgroup.com/en-us/software/vgpnext}, \textit{Bend-Tech}\footnote{https://www.bend-tech.com/}, and \textit{BendPro SIM}\footnote{https://www.currenttech.com/tube-bending-software/bendpro-sim/}), they typically inherit design approaches from industrial manufacturing. These tools provide low-level machine controls rather than giving users tools to \textit{design} parts that can be manufactured with these machines. This makes it difficult for users without technical expertise to design structures for wire bending.

Besides commercial products, researchers have developed software to convert existing 3D models into bend paths~\cite{Huang_Wu_Le_Chen_Lin_Lee_2023,Hsiao_Huang_Chu_2018}, enabling hobbyists to create intricate wire structures without having to deal with machine-specific CAM controls. While powerful, these tools are \textit{conversion tools}, offering users little control and requiring round-trip conversions to edit designs. Moreover, they are agnostic to the constraints of wire bending as a process and the unique material properties of wire structures. Recognizing these challenges, some recent research has proposed more interactive alternatives. WireRoom~\cite{Yang_Xu_Fu_Huang_2021}, for example, approximates 3D geometry using wire bending, but offers users variations to interactively tune the design to accommodate variations. This provides increased control over the aesthetic conversion of design. However, to date, there are no Computer-Aided Design (CAD) tools specifically designed to support wire bending and take advantage of machine and material properties.

We present \textit{Y-AR}, a \textit{fabrication-aware design}~\cite{Bermano:2017:SOT} tool for 3D wire bending in mixed reality, which enables designers to create structures optimized for CNC wire bending. \textit{Y-AR} overlays 3D structures onto users' physical environments in mixed reality, allowing design of objects that fit within the designer's physical environment (such as the mounts, connectors, and braces shown in Figure~\ref{fig:demo object}). We explore the design space of connectors, mounts, and hooks based on wire bending. \textit{Y-AR} is composed of user-facing features to span most of this design space. \textit{Y-AR} leverages the elastic properties of wires to create mechanisms that apply force to hold objects in place. Once the design is ready to manufacture, \textit{Y-AR} lets users export machine instructions for CNC wire benders and generates assembly instructions for users. For fine-tuning, users can still adjust the results in the machine-specific CAM software. Besides guided design through our derived primitives, \textit{Y-AR} implements basic free-form tools (where users draw structures in mid-air) by modeling a custom marble track. 

We found that combining gesture-based interaction with fabrication-aware design principles allowed novice users to create functional wire connectors, even when using coarse XR-based input. We conducted a usability evaluation (n=12) of \textit{Y-AR}, in which participants designed and fabricated a functional bottle holder using \textit{Y-AR}.

%% file: src/01_RELATEDWORK.tex
\section{RELATED WORK}

This work builds on prior research on wire bending, fabrication-aware design, and computer-aided design in extended reality. 

\subsection{Wire Bending}

Research on wire bending has focused on both aesthetic and functional applications. For aesthetic approaches, \textit{WrapIt} created jigs to speed up manual wire bending for jewelry based on 2D patterns~\cite{10.1145/2816795.2818118}. \textit{Fabricable 3D Wire Art} similarly generates jigs, however, they are based on 3D models and images, enabling more complex designs~\cite{Tojo2024Wireart}. With the advent of CNC wire bending machines, there has been an increased focus on advanced wire bending, which involves converting 2D line drawings~\cite {Hsiao_Huang_Chu_2018} or 3D models~\cite{Huang_Wu_Le_Chen_Lin_Lee_2023} to machine instructions for wire bending. \citet{GeneratingVirtualWireSculpturalArtfrom3DModels} propose an algorithm to create 3D wire art with the minimum number of continuous wires. In terms of user interfaces for generating 3D wire art, \textit{WireRoom} offers an interface that provides multiple options for generating wire bending paths~\cite{Yang_Xu_Fu_Huang_2021}, allowing users to select their preferred result. \textit{PM4WireArt} offers users more freedom through a parametric modeling interface designed for conceptual wire art design~\cite{PM4WireArt}. They treat wire as rigid rods assembled into structures, resembling welded frameworks. This approach facilitates the creation of complex designs, highlighting a different use of wire in structural art.

Wire bending has also been studied for its structural and mechanical potential in functional applications. For instance, \textit{Bend-forming} extends the application of wire bending to aerospace engineering~\cite{bhundiya2023bend}. Their work involves generating truss structures and exploring the concept of deploying wire benders in space to construct large-scale, static structures. 

More down-to-earth, \textit{WireFab} generates internal skeletons for existing 3D models, serving as static frameworks with some movement flexibility~\cite{10.1145/3025453.3025619}. 

Finally, \textit{Bend-it} bends kinematic structures by taking key frames of a virtual structure designed by wires as input and generating kinematic mechanisms to animate between these frames, utilizing the springiness and dynamic material properties of the wire~\cite{BENDIT}. 

These fabrication techniques are based on the conversion of 3D structures and the modification of pre-defined parametric models. This enables users to make much more interesting wire-bend content; however, it lacks the ability to really take advantage of the material and machine properties offered by wire bending. In this paper, we argue that a dedicated CAD environment with primitives centered on the relevant machine and material properties will facilitate more accessible workflows and create fabricated results that leverage the springiness and tolerance of wire structures to remain functional despite coarse input or fabrication inaccuracies.

\subsection{Fabrication-Aware Design}

\textit{Fabrication-aware design} is a branch of research in computer graphics that considers the constraints and possibilities of fabrication processes in design environments. This approach helps ensure that the resulting artifacts are manufacturable and functional throughout the design process. \textit{Fabrication-aware design} informs a range of computational fabrication techniques, including laser cutting, 3D printing, and other process-specific workflows.

In the context of laser cutting, \textit{kyub} is a 3D modeling environment that allows users to design models for laser cutting in 3D, while the software handles the conversion to 2D plates by placing joints for the specific material and machine in use~\cite{baudisch_kyub_2019}. \textit{FastForce} safeguards the structural integrity of such laser cut models~\cite{abdullah_fastforce_2021} and \textit{FoolproofJoint} adjusts resulting joints to be optimal for assembly~\cite{park_foolproofjoint_2022}. \textit{HingeCore} leverages both laser cutting and specific material properties to create models that are aesthetically pleasing and easy to assemble~\cite{abdullah_hingecore_2022}. \textit{Fabricaide} allows users to preview fabrication process constraints (such as issues in nesting or joint layout) during the design phase~\cite{Sethapakdi2021}. \textit{FlatFab} similarly supports fabrication-aware design for laser cutting  by focusing on a specific assembly technique of intersecting planar sections~\cite{10.1145/2642918.2647388}. 

In the realm of 3D printing, \textit{fabrication-aware design} tools have also made significant strides. \textit{Bridging the gap} introduces a method to efficiently design stable supports using bridges and pillars~\cite{Bridging_the_gap}. \textit{Stress relief} detects structural weaknesses and provides users with tools to adjust their model accordingly~\cite{Stress_relief}. To increase print efficiency, \textit{WirePrint} offers a rapid prototyping method by generating wireframe structures from existing 3D models~\cite{WirePrint}, and \textit{PackMerger} optimizes the segmentation and packing of printed parts to save material and print time~\cite{vanek2014packmerger}. Others have introduced material-specific properties to fabrication-aware design. For instance, \citet{wang2015saliency} reduces fabrication time while maintaining visual quality through adaptive slicing and segmentation techniques. \citet{Bickel_Bernd} employ data-driven approaches to develop complex, heterogeneous materials with customized deformation properties. \citet{Schumacher_Christian} design micro-structures to achieve variable elasticity using a single stiff material, expanding supported design applications.

There has also been progress in \textit{fabrication-aware design} for textile design and other fabrication methods. In textile design, for example, \textit{Plushie} allows non-professional users to create plush toys by generating 2D patterns from 3D models~\cite{Plushie}. For inflatable structures, \citet{Designing_inflatable_structures} developed a tool that optimizes flat panel patterns, ensuring accurate design through physics-based simulations. In the project \textit{Beyond developable}, researchers explore the use of auxetic materials and conformal geometry to create doubly-curved surfaces from flat pieces~\cite{Beyond_developable}.

Wire bending remains underexplored in terms of \textit{fabrication-aware design}. \citet{Computational_design_of_stable_planar-rod_structures} focus on creating structurally stable wire sculptures rather than serving as a design tool. \textit{Wire mesh design} explores the broader design potential of wire-based materials~\cite{Wire_mesh_design}. Specialized tools for wire bending could expand the design space to include both aesthetic and functional artifacts, supporting connectors, mounts, and mechanisms that are tailored to the wire’s unique material and machine constraints.

\begin{figure*}[t] 
  \centering
   \includegraphics[width=\linewidth]{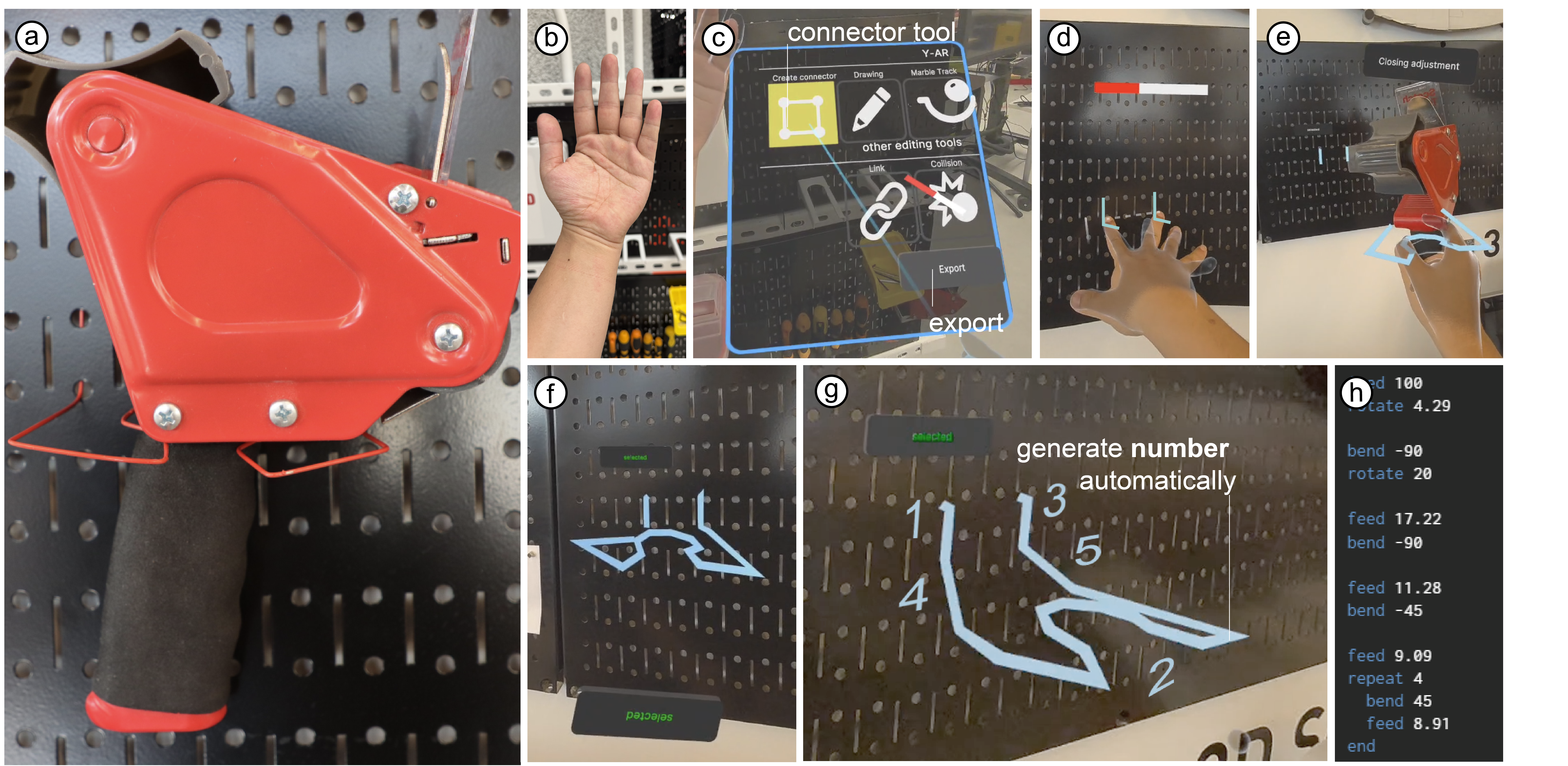}
    \caption{A walk-through of creating a mount for a Scotch tape extruder using \textit{Y-AR}. (a) Final fabricated result on a pegboard. (b) The AR menu launches when the user’s palm is facing towards the user. (c) The user clicks to activate the \textit{Connector Tool}.  (d) The connector tool lets users specify the form of connection using hand gestures. Here, they point their fingers to the pegboard, imitating how a connector would fit into the pegboard--\textit{Y-AR} generates a connector scaled based on a hand gesture. (e) The user activates the connector tool again and holds the extruder to define a handle-fitting mount. (f) The user then reopens the menu to activate the Link Tool; selection labels appear on each wire structure, and the user selects two labels to automatically link them together into a complete wire structure. (g)\textit{Y-AR} generates stepwise assembly instructions displayed in-situ, and (h) exports machine instructions. }
  \label{fig:walkthrough}
\end{figure*}

\subsection{Computer-Aided Design in Extended Reality}

Extended Reality (XR) CAD technologies offer several advantages for digital fabrication. Tools like \textit{pARam}, allow in-situ previewing of designs~\cite{10.1145/3613904.3642083}, \textit{Mix\&Match} enables modeling-free personal fabrication with real-world dimensions~\cite{10.1145/3313831.3376839}, and \textit{MixFab} integrates virtual CAD environments with physical objects to create a mixed-reality workspace that lowers barriers to adoption~\cite{10.1145/2556288.2557090}. Additionally, commercial products such as \textit{Gravity Sketch}\footnote{\url{https://www.gravitysketch.com/}} demonstrate the practical advantage of XR CAD in design, allowing for quick creation of 3D models, even if they lack the precision of traditional CAD tools. Microsoft's \textit{Maquette}\footnote{\url{https://store.steampowered.com/app/967490/Microsoft_Maquette/}}, facilitates spatial prototyping by combining tools like surface snapping, hull brushing, and bounding manipulation for working with 3D content.

XR also presents challenges. One major issue is the difficulty in accurately obtaining environmental information, as users often interact in mid-air, resulting in misalignment between virtual and physical environments. Even though researcher has begun exploring ways to align virtual environments with the real world, such as \textit{SnapToReality}, which proposes a semi-automatic alignment strategy for registering virtual and real spaces, their work relies on scene content and environmental constraints (e.g., edges and surfaces) to anchor virtual content~\cite{10.1145/2858036.2858250}. Another approach, presented by \citet{fotouhi2020reflective} enabling users to visualize different perspectives while actively adjusting the pose. 

Other challenge lies in tracking or gesture detection. Efforts to define hand gestures face limitations, such as high similarity among gestures leading to recognition errors and limited available information~\cite {10.5898/JHRI.5.2.Zhou}. Some solutions, such as high-quality tracking devices, exist but are often prohibitively expensive and inaccessible for everyday home use.

\textit{Y-AR} leverages the strengths of XR CAD, such as the ease of obtaining real-world dimensions, while mitigating its shortcomings. By combining XR CAD with wire as a material, we take advantage of the wire’s springy nature to design mechanisms that tolerate inaccuracies in design, measurement, and fabrication, conceptually similar to how \textit{SpringFit}~\cite{roumen_springfit_2019} overcomes inaccuracies of laser cutting machines. This approach allows us to create adaptable mechanisms in \textit{Y-AR} that maintain stability and functionality despite misalignment, addressing limitations related to precise alignment.

%% file: src/02_USERINTERFACE.tex
\section{Walk-through of \textit{Y-AR}}

Figure~\ref{fig:walkthrough} shows a walk-through of a user creating a mount for a Scotch tape dispenser on a pegboard (Figure~\ref{fig:walkthrough}). The user begins by putting on a Meta Quest 3 headset and invoking the \textit{Y-AR} system. Raising their palm brings up the AR menu, where they select the Connector Tool. To create a pegboard connector, the user points their fingers at the pegboard, and \textit{Y-AR} interprets this gesture to size and position a connector appropriately. Next, holding the tape dispenser itself, the user defines a mount that fits the handle’s size. After placing these connectors, the user switches to the Link Tool to connect them into a unified wire structure.

After design, users export their structure using the export button shown in Figure~\ref{fig:walkthrough}b. Export generates machine instructions for the wire bender (in CSV format, readable by most typical machines), as well as assembly instructions for the user in AR(shown in the Figure~\ref{fig:walkthrough}f). \textit{Y-AR} aims to create structures from a single wire; however, when infeasible \textit{Y-AR }exports multiple wires and provides instructions in AR, which guide users through the process of assembling the wires into a 3D structure, as shown in Figure~\ref{fig:walkthrough}g. To rigidly connect wire segments, we use a Crimping Tool Kit\footnote{\url{https://www.amazon.com/dp/B0C7V6K4X3}} (other similar kits will also work). This allows for secure and precise joining of wires by crimping connectors onto the ends of the wire segments.

\section{contributions, benefits, and limitations}

We articulate three core contributions that collectively advance fabrication-aware design for wire bending:

\begin{enumerate}
\item~We systematically explore the design space of connectors made with wire bending, through which we identify technical guidelines for functional wire bending. 

\item~We contribute a method for modeling and fabricating flexible bent-wire connectors that leverage material stiffness and elasticity to compensate for coarse mixed reality inputs. We validated this approach through a usability study with 12 participants. 

\item~We demonstrate the first fabrication-aware design tool for wire bending, going beyond merely aesthetic considerations or \textit{conversion to wire bending} from other CAD environments.
\end{enumerate}


Limitations include that in this paper, we demonstrate examples using one specific material used for wire bending (1.6mm aluminum). Although the ideas and methods translate to other materials, reconfiguration of some tools in \textit{Y-AR} would be required (indicated in the relative paper sections). \textit{Y-AR} is a demonstration of what a fabrication-aware wire-bending tool could enable. We open-source the code base and invite the community to continue expanding the vocabulary of tools for wire bending.

%% file: src/03_DESIGNSPACE.tex
\section{Design Space}

\begin{figure}[h]
  \centering
   \includegraphics[width=1\linewidth]{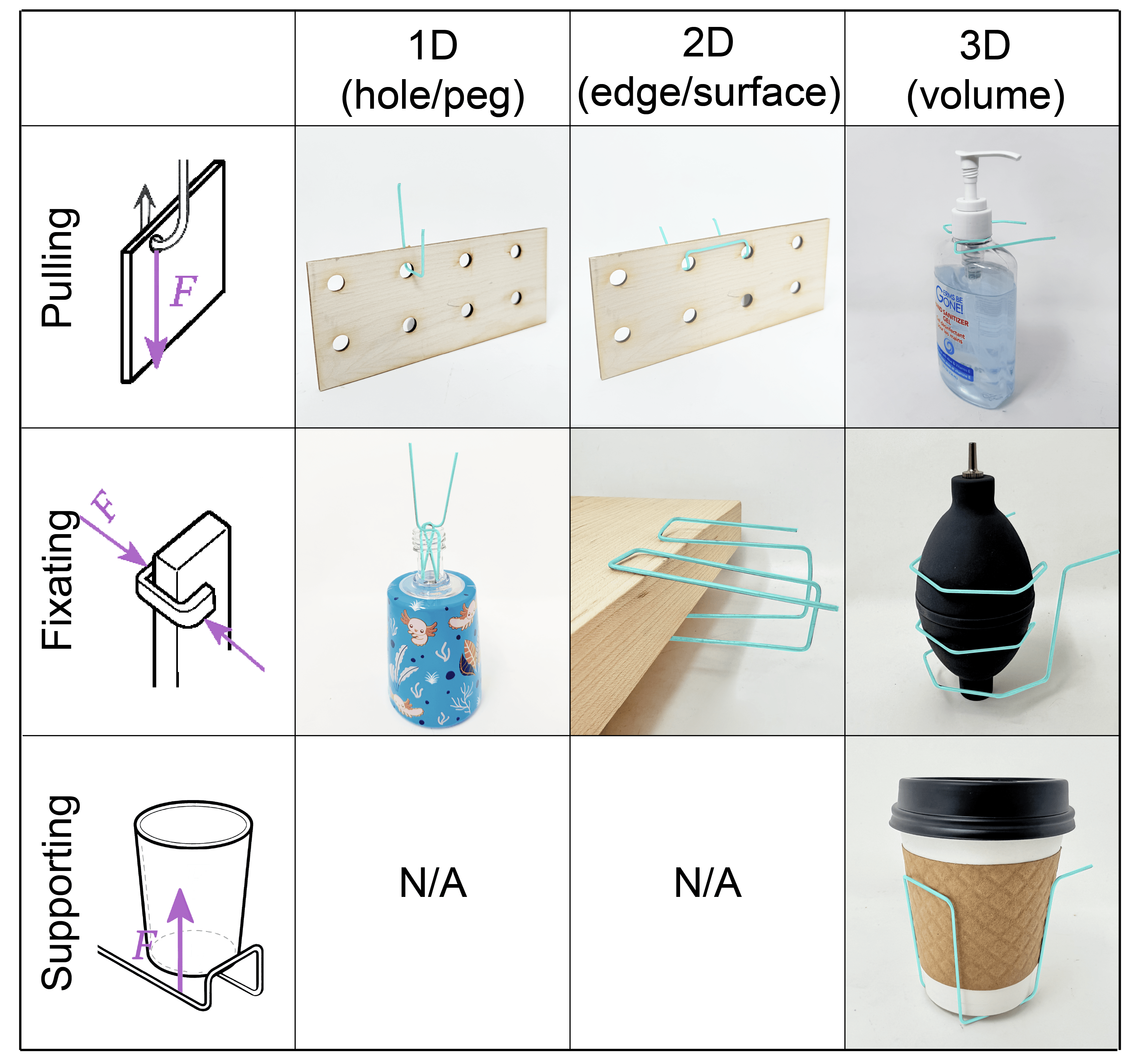}
  \caption{
    Design space of wire structures, organized by the geometry of the target object (rows) and the type of physical engagement (columns). Each cell presents a representative example illustrating how a wire structure can connect to, support, or secure an object in a given configuration.  }
  \label{fig:design-space}
\end{figure}

We define the design space of wire-bent connectors along two key dimensions: the geometry of the target object (1D (holes), 2D (edges), and 3D (volumes)), and how forces are exchanged between connector and object (object pulling on the connector, the connector fixing the object in place, or the connector supporting the weight of the object). 

While artistic wire-bent structures can exist independently, such as in free-form sculptures or spatial frames, their most compelling quality from an engineering and design perspective is their ability to interact with other physical objects. The spring-like behavior of wire enables adaptable attachments that can grip, wrap, or support objects without requiring high-precision modeling or rigid alignment. This makes wire an ideal material for creating fast, robust adaptations in physical contexts. Here, we lay out the design space of wire-bent springs as connectors. Figure~\ref{fig:design-space} illustrates this space.

The rows of the design space correspond to the geometric features of the interface between the object and its connector. 
\begin{itemize}
    \item \textbf{1D:} The wire interacts with a single point or hole, such as inserting into a peg or pulling through a hole.
    \item \textbf{2D:} The wire attaches to an edge or surface, like clipping to the edge of a table.
    \item \textbf{3D:} The wire engages with the object’s volume, wrapping around it or forming an enclosure.
\end{itemize}

The columns of the design space describe how forces are exchanged between the object and the structure:
\begin{itemize}
    \item \textbf{Object pulling on connector}: such as hooks, loops, or closed eyelets.
    \item \textbf{Connector fixating object in place}: such as clamps, paperclip, or toggle lock.
    \item \textbf{Connector supporting the weight of the object}: The object applies force on the wire-bent structure from the bottom, such as cantilever arms and holders. 
\end{itemize}

\subsection{Translation of design space to design patterns}
As we explore the design space and corresponding connectors, we derive specific design guidelines that form the basis of implementation for \textit{Y-AR} and also more broadly inform the design of effective CNC wire-bending.

\subsubsection{Combining connectors}
Connectors, by design, reduce the degrees of freedom between the different objects they connect. The specific way each example in the design space accomplishes this will help negotiate trade-offs between different types of connectors. These approaches are not independent of one another and can be combined at will. For example, when a passive object applies a force to the connector, this may depend on the angle at which the object is held relative to gravity, whereas when the connector applies a force to the object directly, this dependency is much less profound.

\begin{figure}[h]
  \centering
  \includegraphics[width=1\linewidth]{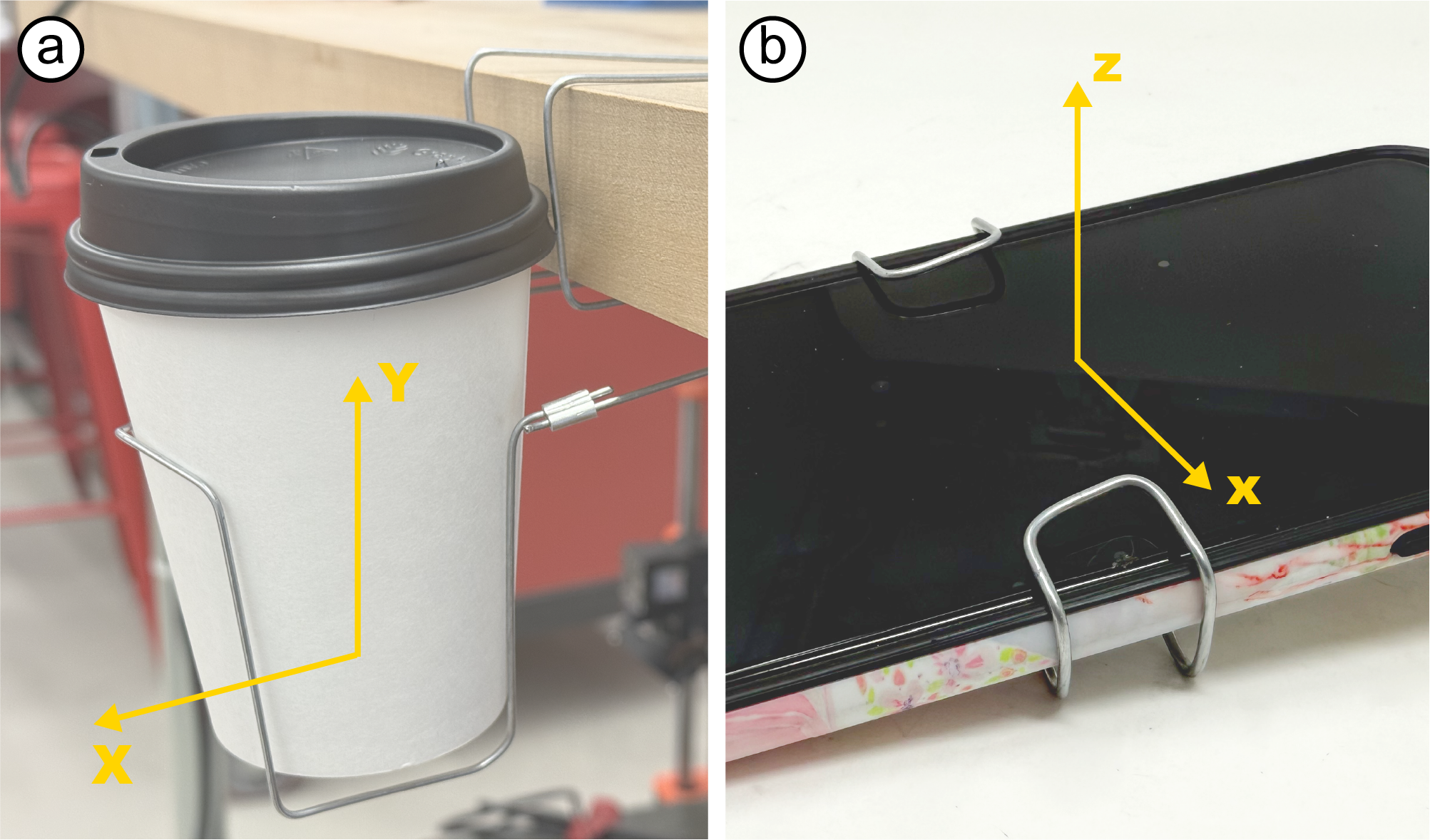}
  \caption{Examples of wire structures that constrain multiple degrees of freedom. (a) A cup holder that combines bottom support with side compression. (b) A phone clamp that applies force from both the x and z directions.}
  \label{fig:3DOF}
\end{figure}

We demonstrate this principle through two examples shown in Figure~\ref{fig:3DOF} a) the cup holder that supports it from below while also pressing in from the sides takes advantage of two connector principles simultaneously, and b) a phone clamp that applies force from both the x and z directions. 

\subsubsection{Fabrication Constraints}
Wire bending is a unique fabrication technique, while versatile and powerful it also introduces its own set of specific constraints when fabricating. We characterize these and demonstrate how we circumvent limitations while maximizing design freedom.

\begin{figure}[b]
  \centering
  \includegraphics[width=1\linewidth]{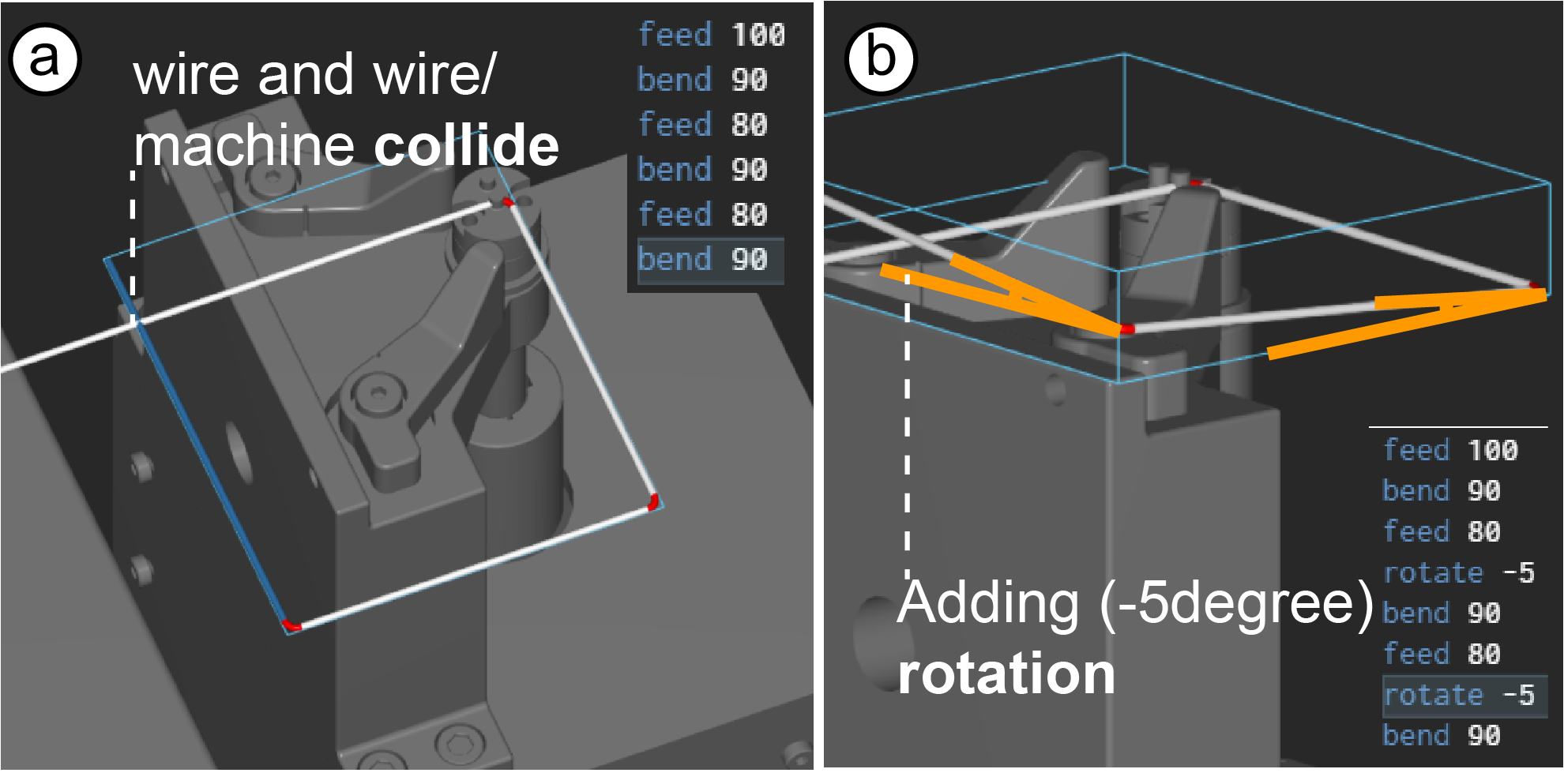}
  \caption{(a) A planar structure design, where multiple bends can lead to collisions. (b) A demonstrates of our strategy:  rotations at the nodes to create non-planar bends, significantly reducing collisions.}
  \label{fig:non-planar}
\end{figure}

\paragraph{Avoid planar bends}. As the complexity of wire structures increases, collision avoidance becomes increasingly crucial (e.g., three consecutive 90-degree bends); the wire may collide with the machine or the extruded wire itself, leading to wire-bending failure. By introducing small rotations at the nodes within planar structures, as shown in Figure \ref{fig:non-planar}, reduces the risk of collisions in the plane. 

\paragraph{Minimum bending angle to overcome plastic deformation.} To plastically deform wires requires overcoming the elastic regime of the material. To overcome this, we identify a minimum required bending angle and over-extend bends where needed. Details of this evaluation are included in the technical evaluation of this paper.

\paragraph{Linking connectors.} 
Structurally, the best way to connect connectors is with at least three non-planar wires, providing rigidity by constraining all degrees of freedom. However, given the linear nature of wire-bending, having a single start and end-point reduces design complexity, enabling fast fabrication while occasionally requiring additional support in one degree of freedom.

\begin{figure}[h]
\centering
\includegraphics[width=1\linewidth]{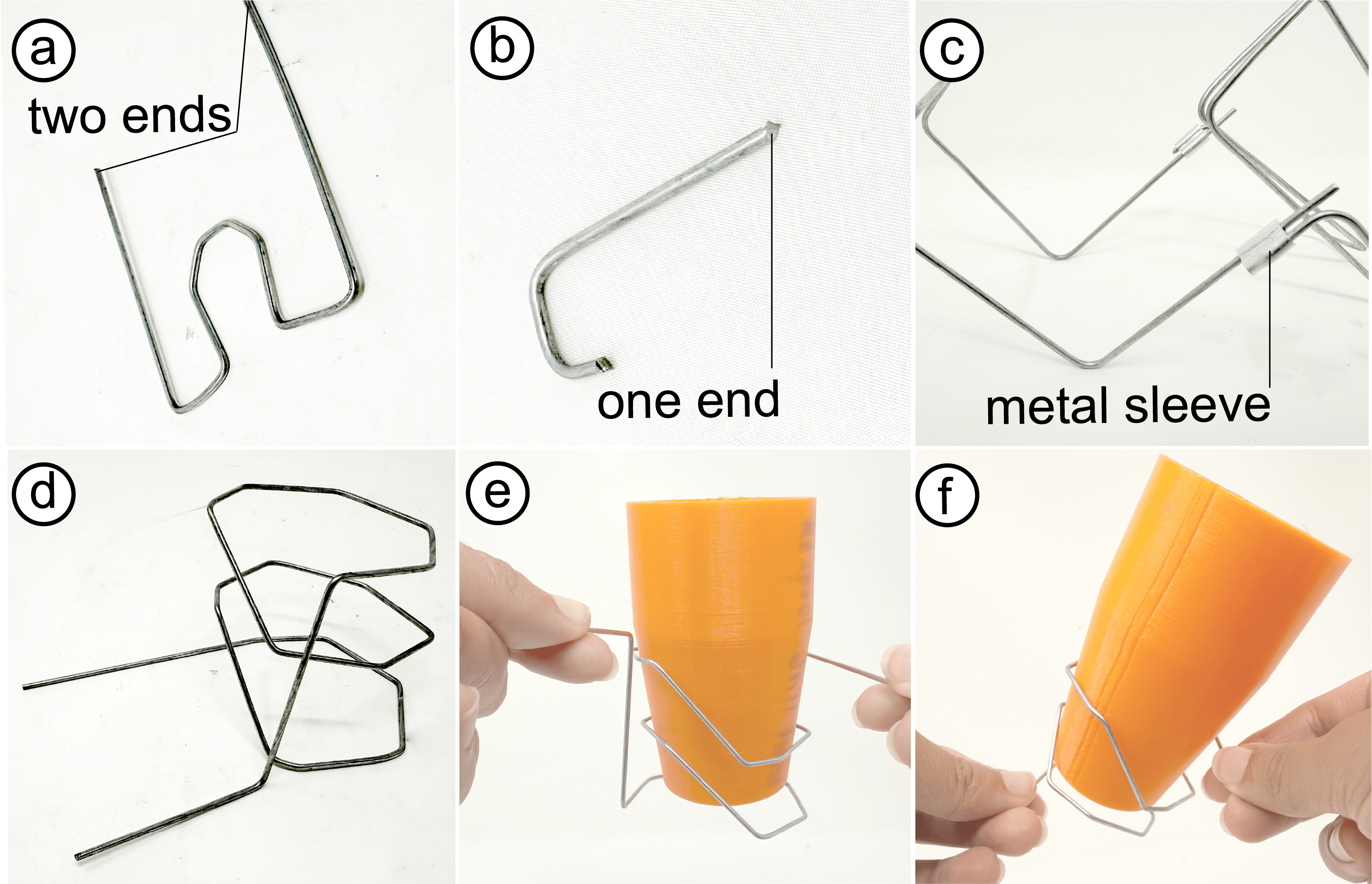}
\caption{(a) A connector with two endpoints joined to other segments. (b) A hook connector with a single connection point. (c) Example assembled structure. (d) A structure with endpoints at equal heights, enabling correct orientation. (e) Example showing how the structure securely holds an object in the intended orientation. (f) An example where misaligned heights cause the orientation of the held object to change.}
\label{fig:connection-strategies}
\end{figure}

\textit{Y-AR} encourages designing structures using as few wire segments as possible to reduce fabrication effort and simplify assembly, ideally forming each functional connector from a single wire with two endpoints (Fig.~\ref{fig:connection-strategies} a). These wire ends are joined to other segments or structures using metal sleeves and a crimping tool kit\footnote{\url{https://www.amazon.com/dp/B0C7V6K4X3}}, forming a complete assembly (Fig.\ref{fig:connection-strategies} c).

This simplification requires careful attention: many wire structures consist of connectors with two endpoints at approximately equal relative heights, which simplifies assembly but introduces sensitivity to orientation. When joining two wire endpoints using crimping (Fig.\ref{fig:connection-strategies}c), designers must consider not only the position but also the relative height and orientation of the connector with respect to the object being gripped. Even if endpoints are correctly positioned, a misalignment between the connector’s grasping direction and the intended load direction can cause the assembled structure to twist or fail to hold the object as intended. For example, the structure shown in Fig.~\ref {fig:connection-strategies} d, Fig.\ref{fig:connection-strategies}e shows a structure where the connector properly holds an object in the correct orientation, while Fig.~\ref{fig:connection-strategies}f shows the same structure with a height misalignment, resulting in a rotated or unstable grasp.

In some designs, such as hook-based connectors, only one end actively engages with another structure while the remaining free end must be positioned thoughtfully to maintain overall balance (Fig.\ref{fig:connection-strategies}b). An example assembled structure, showing how crimping is used to join wire segments securely (Fig.\ref{fig:connection-strategies}d).

\begin{figure}[b]
  \centering
  \includegraphics[width=1\linewidth]{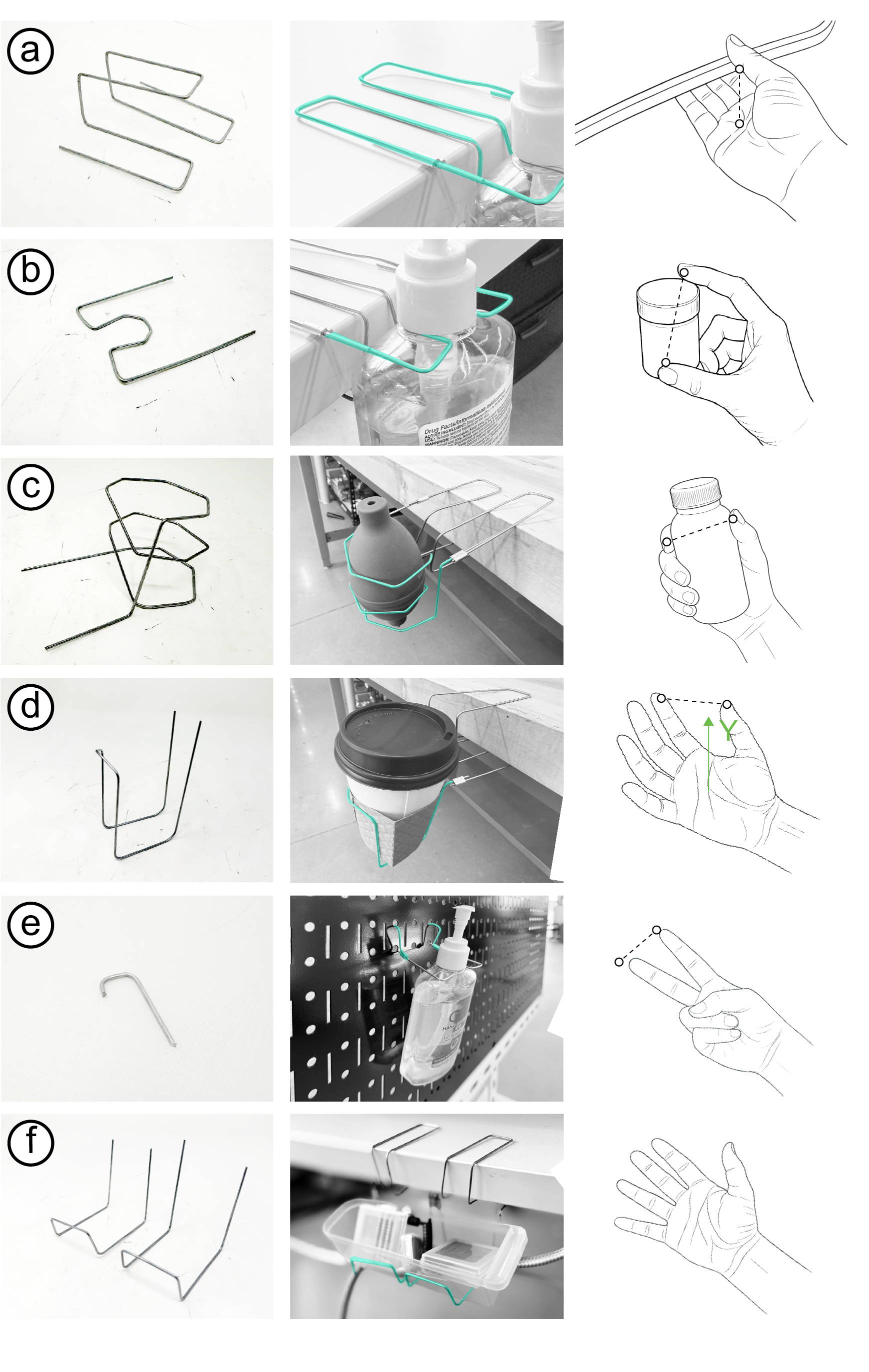}
  \caption{Illustrates gesture-to-structure mapping for six common connector types. Users can immediately modify the generated connector through rotation, translation, and scaling once it appears as a preview in the scene. Dashed lines indicate the distances measured from each gesture. }
  \label{fig: te}
\end{figure}

\section{Design of \textit{Y-AR}}

\textit{Y-AR} adopts a verb-noun interaction model to make actions explicit and easy to discover, helping users understand what they can do (verbs) and what objects they can act upon (nouns). The \textit{Y-AR} interface consists of a top-level menu with various actions, including creating connectors, mid-air drawing, and creating marble tracks. For basic editing, users can link structures, delete components, or move them. To transition to fabrication, it has an export function. In this section, we discuss the design rationale behind these tools.

\subsection{Create Connector Tool}
\begin{figure}[b]
  \centering
  \includegraphics[width=\linewidth]{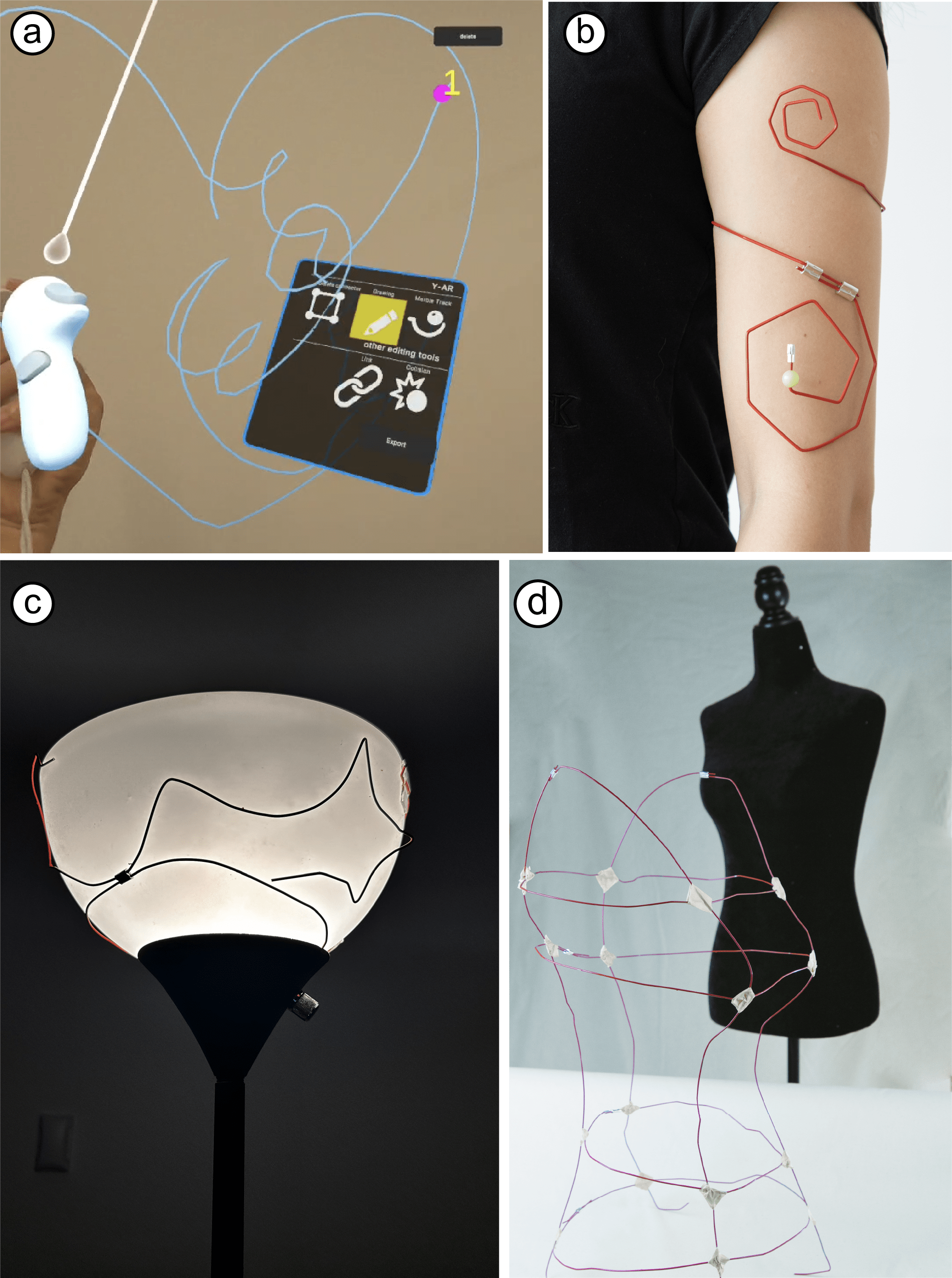}
  \caption{Users can draw in various contexts, such as (a) mid-air, (b) on the human body, (c) on the surfaces of existing objects, like lampshades for decorative purposes, or (d) create mesh structures based on dress forms.}
  \label{fig:Midair}
\end{figure}

\begin{figure*}[t]
  \centering
  \includegraphics[width=\linewidth]{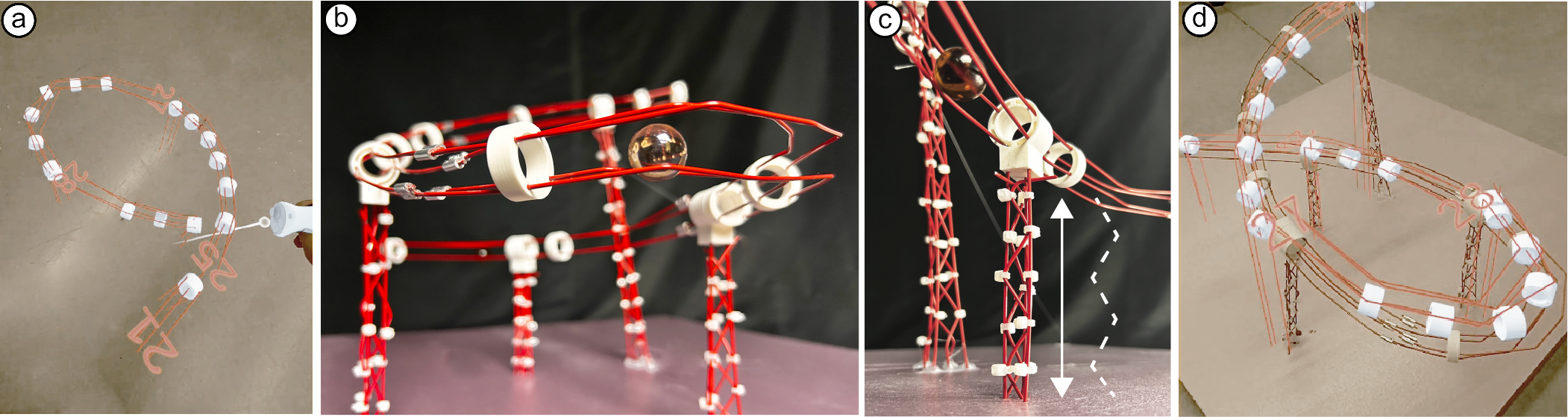}
  \caption{The user draws a marble track in mid-air. (a)~Instead of single lines, \textit{Y-AR} renders a full track. (b)~It also generates supporting structures at the ends of the track. (c)~After completing a stroke, \textit{Y-AR} exports an additional CSV file for reinforced structures between the supports, based on the track's height. (d)~The final structure is assembled after wire bending, following the projected design.}
  \label{fig:marble-track}
\end{figure*}

The Create Connector tool allows designers to create wire-bending structures that connect to existing objects or features within the user's context. The workflow for this tool is as follows:

Users first press the gesture button to enable gesture detection (note that this feature is not "always on" to prevent undesired detection of gestures). They grab the object the way they want to connect, \textit{Y-AR} detects the gesture, and extracts dimensions to scale the connector. For example, dimensions may include the distance between the index finger and thumb to control the size of the generated connector, the hand's orientation, and the palm's position to control placement, depending on the type of gesture. 
The generated connector shows up as a preview of the wire structure. Upon placement, the modify tool is triggered, allowing users to immediately move, rotate, and scale the connector as needed. 

\textit{Y-AR}'s gestures intentionally mimic how a wire structure would physically interact with an object. For instance, a pinching gesture naturally corresponds to a clamping connector; a grab posture resembles wrapping a cylinder. This embodied mapping helps users intuitively specify both the intent and form of the connector during creation.

Below, we provide a more detailed description of how these connector types correspond to their respective gestures. (see Figure \ref{fig: te})

\subsection{Mid-Air Drawing Tool}

The mid-air drawing tool lets users generate paths with more flexibility. As shown in Figure~\ref{fig:Midair}, users can draw in various contexts, such as (a) mid-air, (b) on the human body, (c) on the surfaces of existing objects, like lampshades for decorative purposes, or (d) to create mesh structures based on dress forms. These capabilities enable the creation of unique wire structures, such as wire-wrapped jewelry, which benefit from lightweight connections, high flexibility, and aesthetic customization.

\textit{Y-AR}, users employ the Logitech MX Ink, a pen-based tool for Mixed Reality (MR), to draw lines in mid-air. 

By pressing the button on the MX Ink, users begin drawing; each line ends upon releasing the button. Alternatively, users perform a pinch gesture to start drawing a line, which continues until the gesture ends—eliminating the need for the MX Ink. Users can freely draw lines to model their desired structures directly in 3D space. 

\subsection{Marble Track Tool}
To create more deliberate functional structures, we let users design structures using the \textit{create marble track} tool, demonstrated in Figure~\ref{fig:marble-track}. Users create freehand structures consisting of pre-defined modules. This tool is developed based on the Mid-Air Drawing Tool, and its basic workflow is similar.

When users sketch a line for a marble track, \textit{Y-AR} automatically generates not just a single wire, but four parallel lines to form a robust track. It injects support towers, consisting of vertical truss structures, at the starting point of each drawn wire segment to provide an anchor for stability. Additionally, 3D-printed connectors are placed along the wire's length at every 5 cm to secure the wires together and ensure structural rigidity. Our current marble track modules are optimized for standard 1.6 cm diameter marbles, and the geometric parameters used in track generation (such as minimum bend radius and spacing) follow the same machine-specific simplification logic described in Section~\textit{Simplification Parameters}.

\subsection{Editing tools}

\textit{Y-AR} includes a set of simple editing tools. These tools can be applied to all structures and content in the scene. They consist of a delete tool to remove segments, a modify tool to move, scale, or rotate segments, and a link tool to link individual connectors to one another. Finally, there is an export function that converts the designed wires to machine instructions (CSV format) and assembly instructions (AR preview).

\subsubsection{Simplification Parameters }

Geometric processing steps in \textit{Y-AR} are informed by the physical constraints of the bending machine and material. When simplifying user-drawn lines, we apply smoothing and point reduction based on a minimum bendable angle and a minimum spacing between bends, ensuring that generated geometries remain feasible for fabrication. Additionally, we expose both smoothing strength and minimum point reduction ratio as adjustable settings, allowing users to control the degree of simplification.

\begin{figure}[h]
  \centering
  \includegraphics[width=1\linewidth]{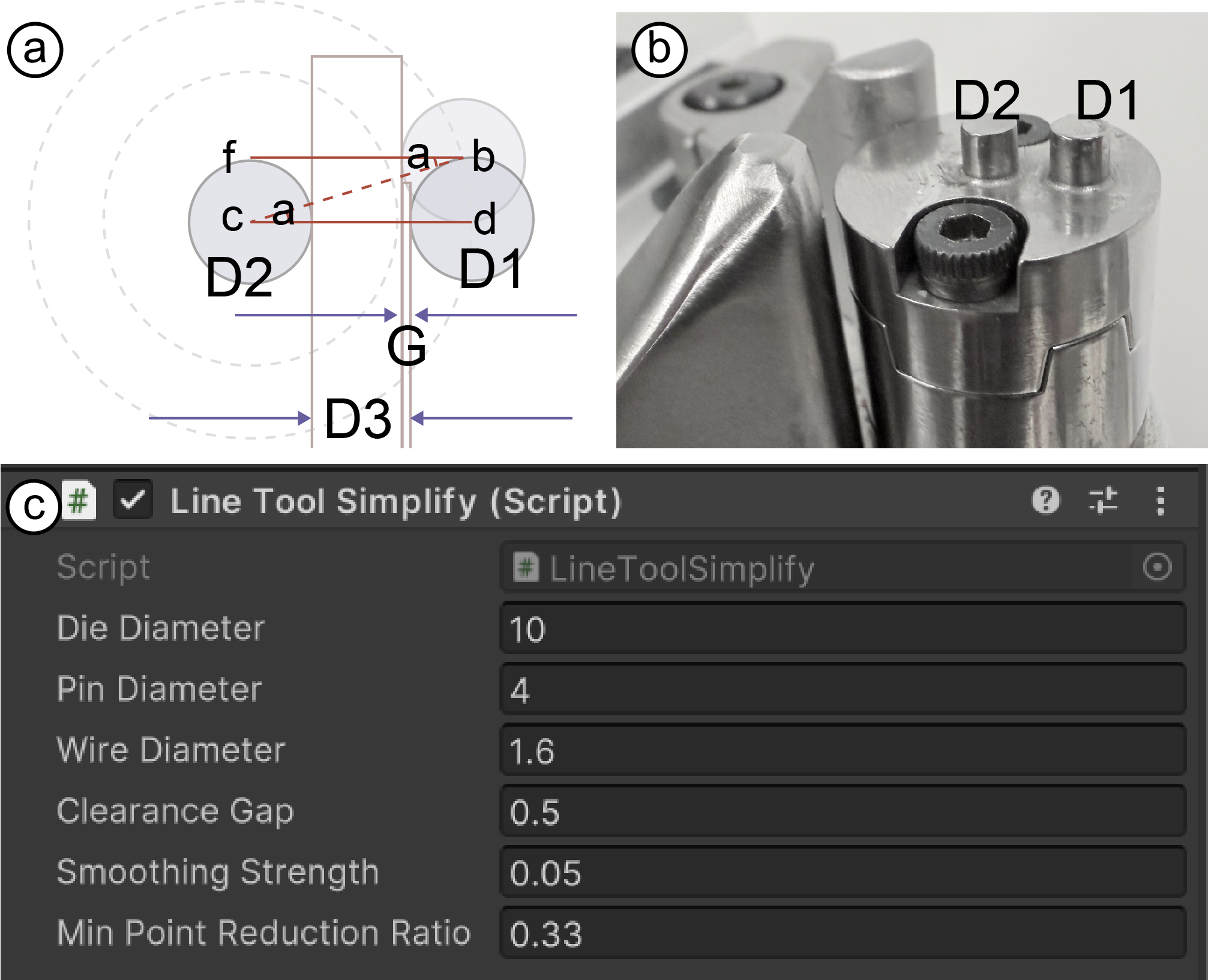}
  \caption{(a) Diagram showing key dimensions used in our geometric analysis: die diameter (D1), pin diameter (D2), wire diameter (D3), and optional clearance gap (G). 
(b) Photograph of the wire bender head used in our experiments. 
(c) Screenshot of Unity interface for adjusting simplification parameters.}
  \label{fig:machine-geometry}
\end{figure}

These parameters are computed from a simple geometric analysis of the bending machine’s head. Specifically, we calculate the minimum distance required between adjacent bends by measuring the die diameter (D1), pin diameter (D2), wire thickness (D3), and an optional clearance gap (G), as shown in Fig.~\ref{fig:machine-geometry}a. The minimum bendable angle is derived from these measurements using the following relations: 

$$
CD = \frac{D1}{2} + \frac{D2}{2} + D3 + G, 
\quad 
CB = \frac{D1}{2} + \frac{D2}{2} + D3,
$$

$$
A = \text{acos}\left( \frac{CD}{CB} \right),
\quad 
\text{Minimum Feed} = CD \times \sin(A).
$$

Fig.~\ref{fig:machine-geometry}b shows a photograph of the wire bender head used in our setup. These measurements ensure that simplification algorithms remain sensitive to the physical limitations of different wire benders and wire diameters. While currently configured for 1.6mm aluminum wire and a specific bending head, these constants can be easily adjusted to accommodate other materials and machines. 

\subsubsection{Collision detection}

\begin{figure}[t]
  \centering
  \includegraphics[width=1\linewidth]{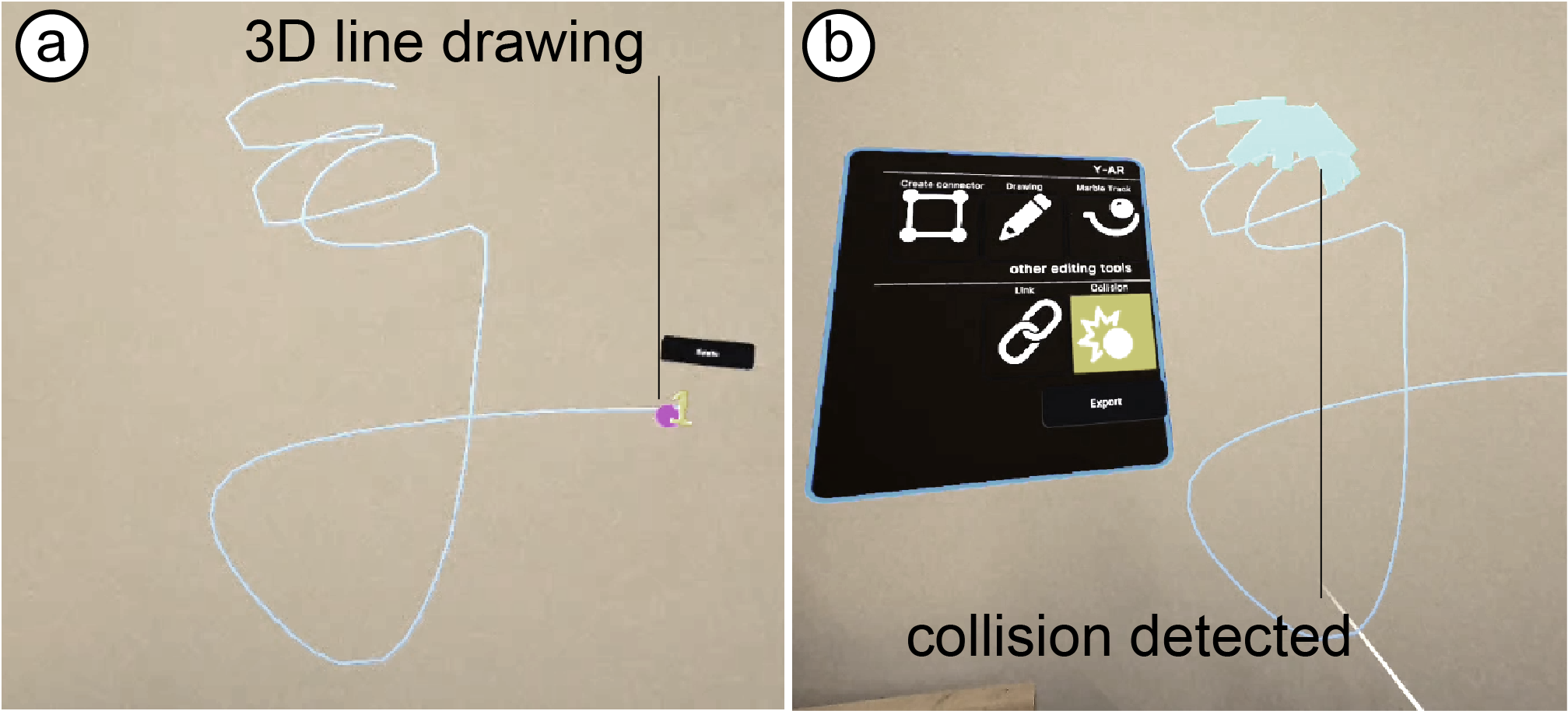}
  \caption{(a) After completing the line drawing, when collision detection is enabled in the user interface, (b) the tool indicates where along the line collisions may occur during the bending process. It does this by generating green markers to highlight the potential collision points.}
  \label{fig: Collision detection}
\end{figure}

After simplification, \textit{Y-AR} analyzes the resulting paths for potential fabrication issues. It sets a minimum threshold angle that the wire bender can fabricate. Since all lines are drawn (as shown in the Fig~\ref{fig: Collision detection} a) by connecting multiple points, the algorithm separates all points and combines every pair of points into a line segment. 

By calculating the angles between two such segment lines, the algorithm determines whether their angle exceeds the threshold. Conflicting segments are flagged, and \textit{Y-AR} visualizes them by drawing short green highlight lines directly between the closest points on the conflicting segments using a distinctive material (as shown in the Fig~\ref{fig: Collision detection} b). This provides an immediate visual indication of where the bending machine may encounter issues. It serves as an informative prompt, nudging users to make manual adjustments. While future work could enable automatic path adjustments or suggest alternatives, such functionality falls outside the scope of this paper.

\subsubsection{Machine instructions and assembly instructions export}

\begin{figure}[b]
  \centering
  \includegraphics[width=1\linewidth]{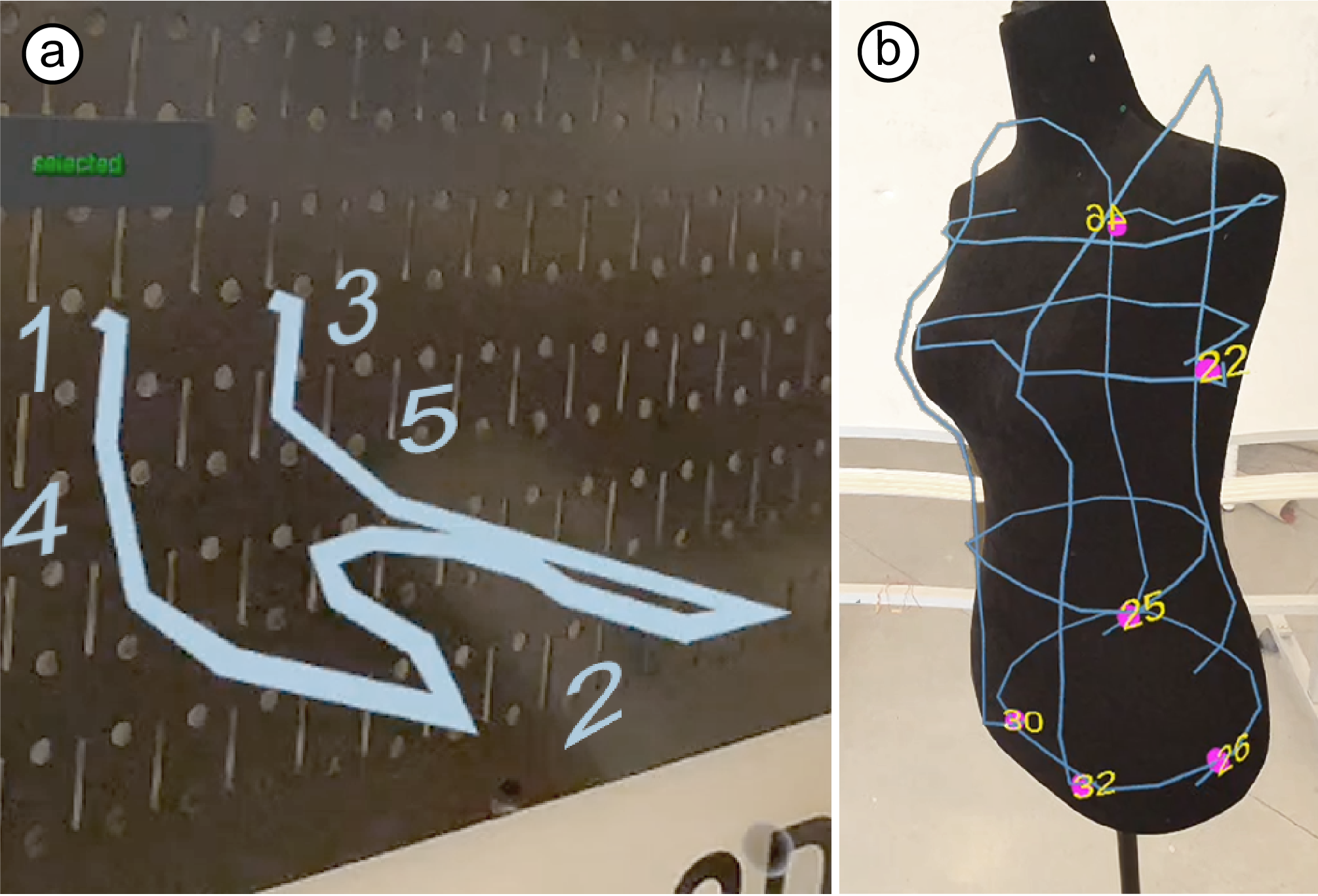}
  \caption{AR preview after export. (a) shows labeled parts from the Connector Tool; (b) shows labeled line segments from the Line Drawing Tool. The numbering matches the exported file names, helping users trace, and organize parts.}
  \label{fig:ar-preview}
\end{figure}

When users finish their design, \textit{Y-AR} exports the wire structures as CSV files and continues to display a preview in AR to assist with assembly.


For connector-type structures, \textit{Y-AR} generates separate CSV files for the connector ends and the wire segments between them. These parts are labeled directly on the virtual objects (Fig.~\ref{fig:ar-preview} a), making it easy to match exported files with assembly steps.  


For structures created using the Line Drawing Tool, the system exports one CSV file per line segment. Each line receives a number label in AR as soon as it is drawn (Fig.~\ref{fig:ar-preview} b), so even complex designs with many lines are easy to trace.  

After export, \textit{Y-AR} keeps the numbered preview visible in AR, so users can clearly see where every part should go and assemble their design accurately.

\subsection{Gesture training}

\begin{figure}[h]
  \centering
  \includegraphics[width=1\linewidth]{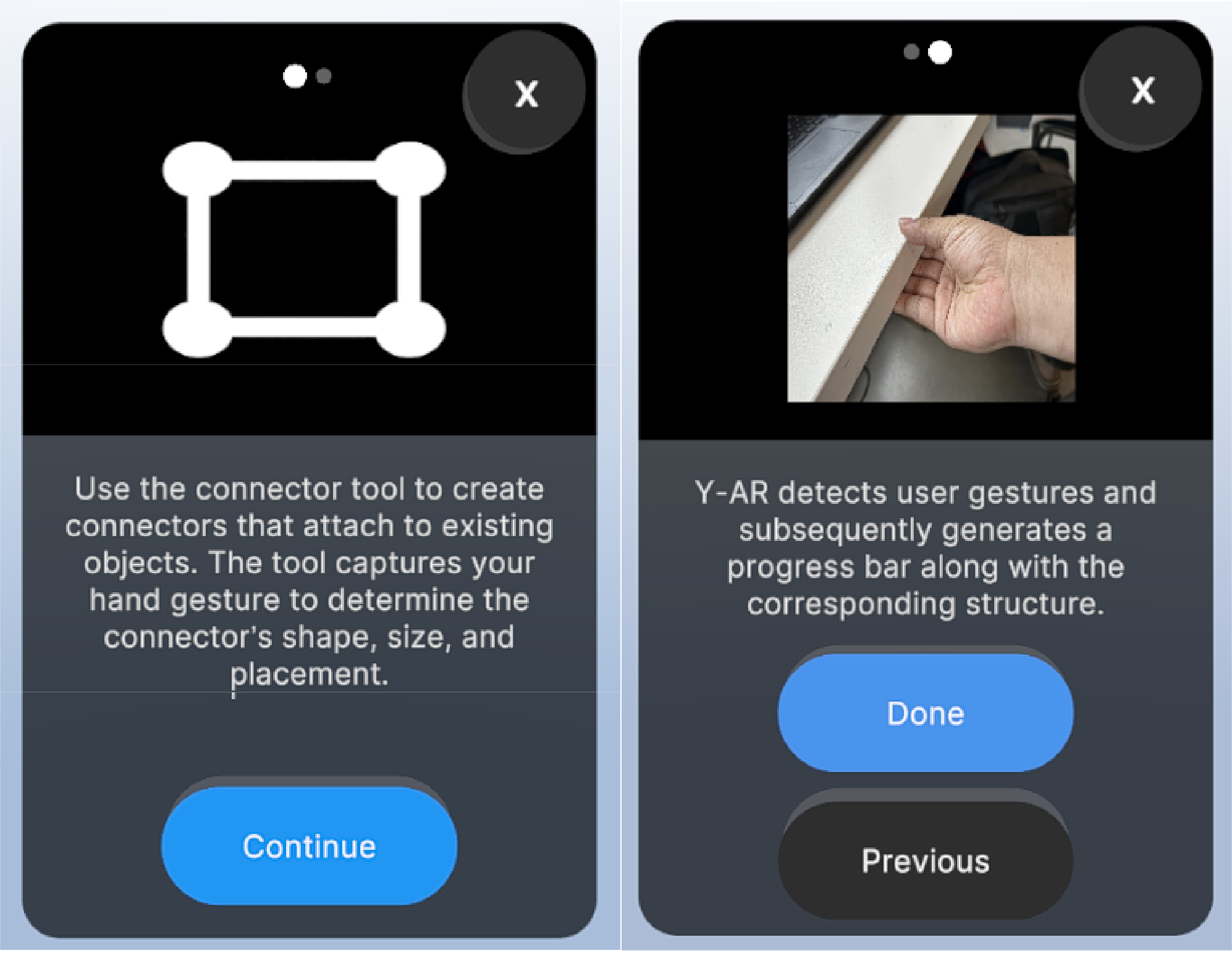}
  \caption{Every tool comes with an onboarding page for first-time users. }
  \label{fig:onboarding}
\end{figure}

Gesture-based interfaces are not the most discoverable~\cite{10.5898/JHRI.5.2.Zhou}. Yet, we believe the intuitive nature of mimicking connector forms with hand gestures offers a clear benefit to end users. In doing so, we build on work by ~\citet{pei2022hand}, which lets users’ hands become virtual objects by imitating the objects themselves. To aid users in adopting these gestures, every tool in \textit{Y-AR} includes an interactive on-boarding mode, as shown in Figure~\ref{fig:onboarding}. This is only triggered upon the first usage of the tool, users can reset them so a different user on the same account can see on-boarding tutorials again.


\section{Implementation}

\textit{Y-AR} is implemented in Unity3D (C\#). It implements 3D and UI interactions via Unity input events and includes customizable 3D UI elements for on-boarding, spatial data collection, and \textit{Passthrough} mode setup. The chosen tech stack, including Unity version 2022.3.8f1, XR Interaction Toolkit (version 2.5.1)\footnote{https://docs.unity3d.com/Packages/com.unity.xr.interaction.toolkit@2.5/manual/}, XR Hands (version 1.4.1)\footnote{https://docs.unity3d.com/Packages/com.unity.xr.hands@1.4/manual/}, XR Core Utilities (version 2.2.3)\footnote{https://docs.unity3d.com/Packages/com.unity.xr.core-utils@2.2/manual/index.html}, and XRC Line Tool (version 0.1.0)\footnote{https://xrcollaboratory.github.io/edu.cornell.xrc.tools.line/}, was selected for its comprehensive support for mixed reality development and ease of integration with various XR features.

\textit{Y-AR} uses the Unity XR Hands package for hand tracking. Gesture recognition translates hand movements into CAD system actions, with real-time feedback provided by the XRC Line Tool, which is used for creating, rendering, and manipulating lines in the project.

\textit{Y-AR} supports mid-air drawing and dynamic placement/scaling of virtual connectors by using XR Interaction Toolkit input events and XR Hands joint positions. We implemented additional logic for recording spatial data and real-time object scaling, which is visualized using the XRC Line Tool. \textit{Y-AR} achieves mid-air drawing by using the controller's position as input, visualizing this with the XRC Line Tool. The system utilizes compatible controllers to track spatial movements, which is then translated into wire-bending paths and exported as drawing data.

The bending paths designed in \textit{Y-AR} are exported as CSV files, which contain a list of coordinates along the path. This format is chosen for its simplicity and compatibility, allowing the files to be easily read and interpreted by wire bending software. This ensures that virtual designs are accurately translated into physical wire forms.

%% file: src/05_USERSTUDY.tex
\section{Technical Evaluation}

We run two technical evaluations to validate the utility of our technique: the structural integrity of our basic connectors and the accuracy of bending angles. These findings inform the implementation of \textit{Y-AR}. In our evaluations, we use 1.6mm aluminum wire, a commonly available material for wire bending. We use this fixed material to characterize the general behavior and efficacy of our technique, given that our contribution is not the calibration of wire-bending machines.

\subsection{Pulling Force Resistance in Spring Wire Connectors}

\textit{Y-AR} leverages the inherent springiness of bent wire structures, which enables secure attachment, even when fabricated with small geometric inaccuracies. To evaluate this property, we conducted a mechanical test using a representative connector designed to hold a paper cup on the edge of a table.
\begin{figure}[h]
  \centering
  \includegraphics[width=\linewidth]{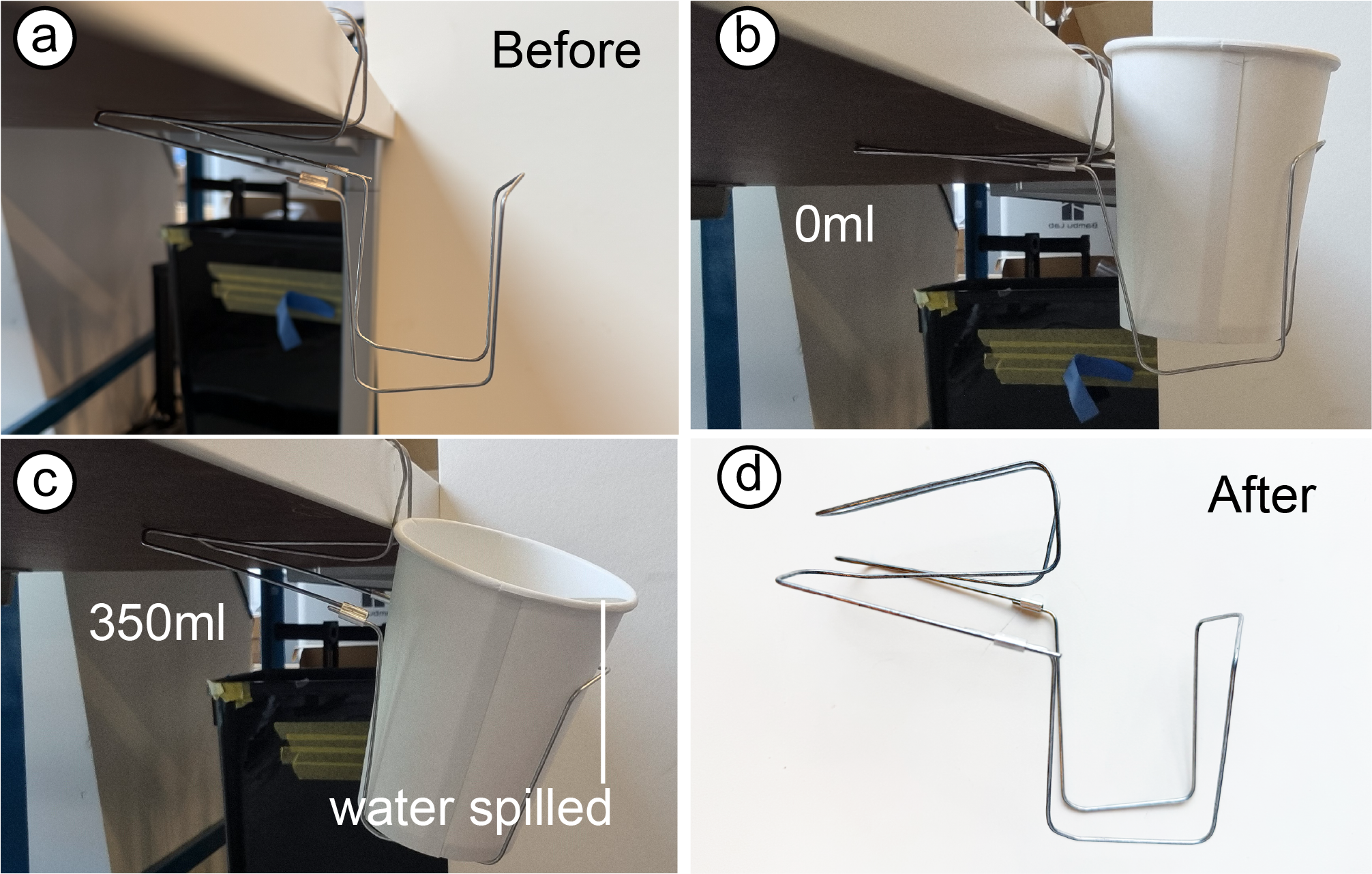}
\caption{Pulling force evaluation of a spring wire connector: (a) the connector configuration before loading; (b) initial loading stage with an empty cup carefully placed; (c) peak load when water begins to spill at 350 ml; (d) connector geometry after test, showing moderate elastic deformation.}
  \label{fig:gripstrength}
\end{figure}

We fabricated the connector using 1.6 mm aluminum wire and incrementally increased the load by filling the cup with water in 50 gram increments. This process simulates an increasing downward pulling force, and we define failure as the point at which the structure can no longer hold the cup or when water starts to spill.

In this test, water spills out when the load reaches 351 grams (350ml of water and the weight of the cup), as shown in Figure~\ref{fig:gripstrength}c. Even as water begins to spill, the structure continues to hold the cup securely. The structure displays only moderate elastic deformation (as shown in the Fig~\ref{fig:gripstrength}d
). The sequence of loading stages illustrates the progressive deformation leading up to this point, as shown in Figure~\ref{fig:gripstrength}b-c. 

We conclude that the connector does fulfill its purpose until the cup is effectively full, to connect heavier objects, further reinforcement or thicker wire will be required. The software indicates an approximation of the weight connectors can hold. Future work could automatically reinforce when that value is exceeded.

\subsection{Assessing Bending Accuracy for Small Angles}

Wire bending requires overcoming both elastic and plastic deformation, as a result of this, simply bending to a certain angle may not suffice as the elastic deformation (or springiness) will make the material bounce back to some extend. Here, we characterize the bending behavior compared to our target angles.

\begin{figure}[h]
  \centering
  \includegraphics[width=1\linewidth]{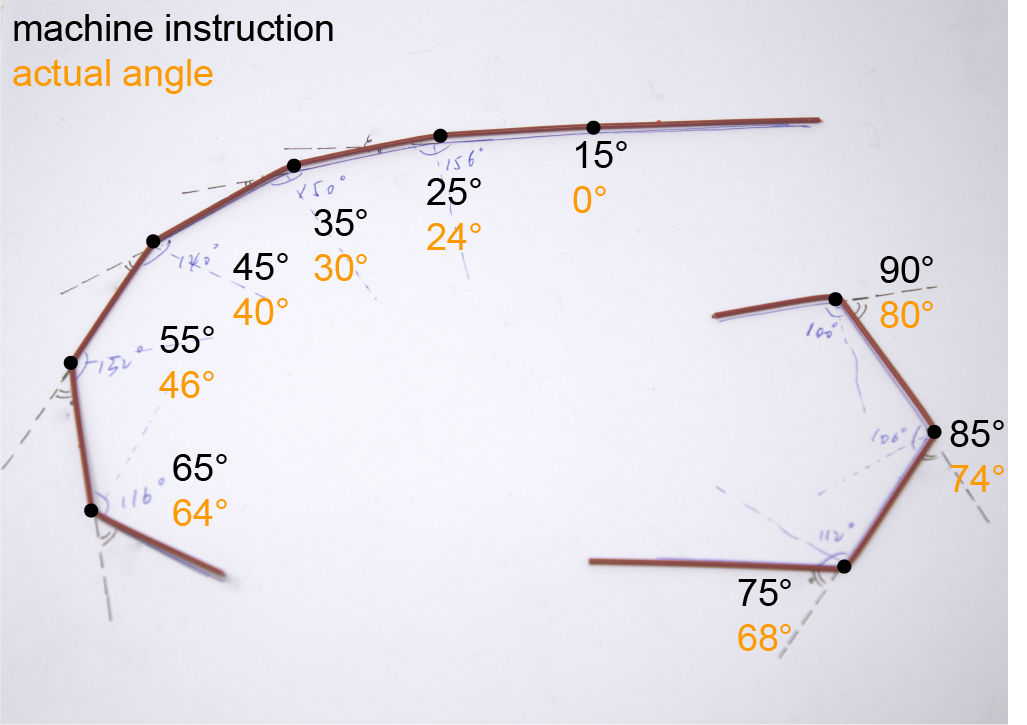}
  \caption{Comparison of machine bending angles (black) with actual bending angles (orange). The results show that when the machine was set to bend at 15 degrees, the actual achieved angle was nearly zero, highlighting the need to avoid specifying angles smaller than 15 degrees in the design.}
  \label{fig:minim}
\end{figure}

We evaluate bending angles from 5$^{\circ}$, 15$^{\circ}$, 25$^{\circ}$, 35$^{\circ}$, 45$^{\circ}$, in 10$^{\circ}$ increments up to 85 degrees and compare to measured angles.We repeated this test five times and observed consistent results, Figure~\ref{fig:minim} shows one representative trial.

One critical result is that for small angles, a significant over-bend is required, a 15$^{\circ}$ of less bend will not achieve plastic deformation. In \textit{Y-AR} this finding is used to avoid short bends with small angles, and if required by design they are accomplished by over-bending past the 15$^{\circ}$ mark. For larger angles some over-bending is still required, but to a lower extend.

\section{User Study}

We recruited participants to design using \textit{Y-AR}. To assess the usability of \textit{Y-AR}, we analyze their reactions and feedback.

\paragraph{Participants.} We recruited 12 participants (9female/3male/0 other; mean age = 21.4, SD = 4.0) from our university (see Table~\ref{tab:expert-participants}). We asked participants how frequently they used CAD tools and XR. P1, P3, P6, P7, P9, P11, and P12 reported prior exposure to wire-bending.

\paragraph{Apparatus.} Participants used a Meta Quest 3s headset, with the Logitech MX INK stylus as input controller. All fabricated wire structures were fabricated on the PensaLabs C64 CNC Wire Bender\footnote{https://www.pensalabs.com/c64}, with 1.6mm aluminum wire.

\paragraph{Training.} Participants received a brief on-boarding session, including headset setup, a built-in tutorial on hand gestures, and a short demonstration video showing key \textit{Y-AR} features and a complete modeling example.



\begin{figure}[h]
  \centering
  \includegraphics[width=1\linewidth]{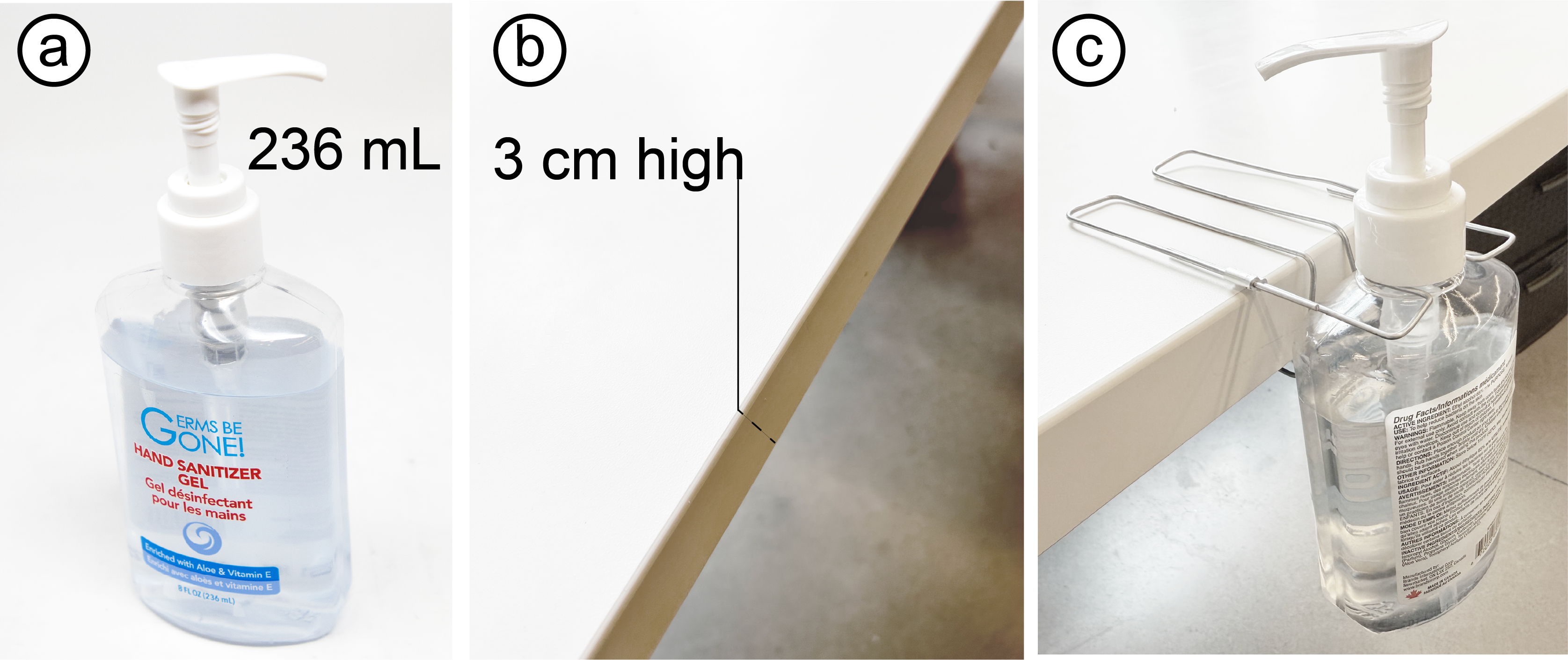}
  \caption{a) The specific bottle, b) table edge, and c) the example result of Task 1.}
  \label{fig:connector-for-bottle-and-table-edge}
\end{figure}

\textbf{Task 1:} Participants were tasked to create a functional object in response to a clearly defined real-world need. Specifically, they were asked to design a wire connector that would attach a standard 236 ml plastic bottle to the edge of a flat table (as shown in Fig.~\ref{fig:connector-for-bottle-and-table-edge}).

\begin{figure*}[h]
\centering
\includegraphics[width=1\linewidth]{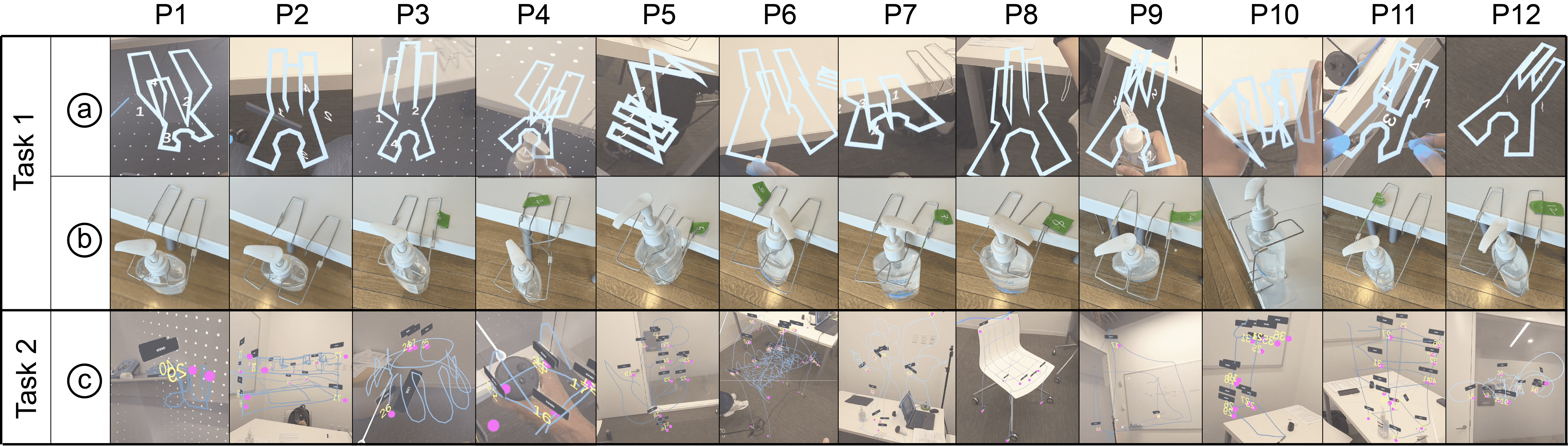}
\caption{Participant outcomes: (a) all Task 1 results; (b) fabricated connectors holding the bottle; (c) samples from Task 2.}
\label{fig:results_tasks}
\end{figure*}

Before designing, participants were encouraged to think about how they might physically secure the bottle to the table edge using their own hands, and then to explore corresponding gestures in \textit{Y-AR} to design their connector.


\textbf{Task 2:} An open ended task using \textit{Y-AR}'s freehand design tool. Participants were encouraged to explore the surrounding environment—such as tables, chairs, shelves, and cables—and construct wire structures that interact with these elements. During the process, participants were encouraged to discuss their design decisions aloud and iteratively model their ideas using \textit{Y-AR}.

\paragraph{Procedure} Each study session lasted approximately 60 minutes and was conducted in a controlled laboratory environment. 

After training, participants proceeded to Task 1, with up to 20 minutes to design a functional wire connector. Upon completing the design, participants filled out a System Usability Scale (SUS) questionnaire reflecting on their experience.

They then proceeded to Task 2 for 15 minutes. At the end of the session, participants took part in a short semi-structured interview to discuss their design decisions, challenges, and overall impressions of \textit{Y-AR}.

\paragraph{Metrics} We collected both subjective and objective metrics:
\begin{itemize}
  \item \textbf{System Usability Scale (SUS):} Participants rated the overall usability of \textit{Y-AR} using the 10-item survey~\cite{brooke_sus_1996}.
  \item \textbf{Mechanical success:} We evaluated whether the fabricated wire structures could securely perform their intended function (e.g., supporting a bottle on a table edge).
  \item \textbf{Semi-structured interviews:} We analyzed qualitative feedback collected from post-task semi-structured interviews.
\end{itemize}

\subsection{Results}

The mean SUS score across all participants was 77.3. All participants completed Task 1 successfully within the time limit.

In mechanical testing, all connectors produced in Task 1 reliably supported the bottle. As shown in Fig.~\ref{fig:results_tasks}, participants generated varied designs that successfully conformed to the target bottle.In Task 2, participants created diverse freeform wire structures. 

\begin{figure}[h]
\centering
\includegraphics[width=1\linewidth]{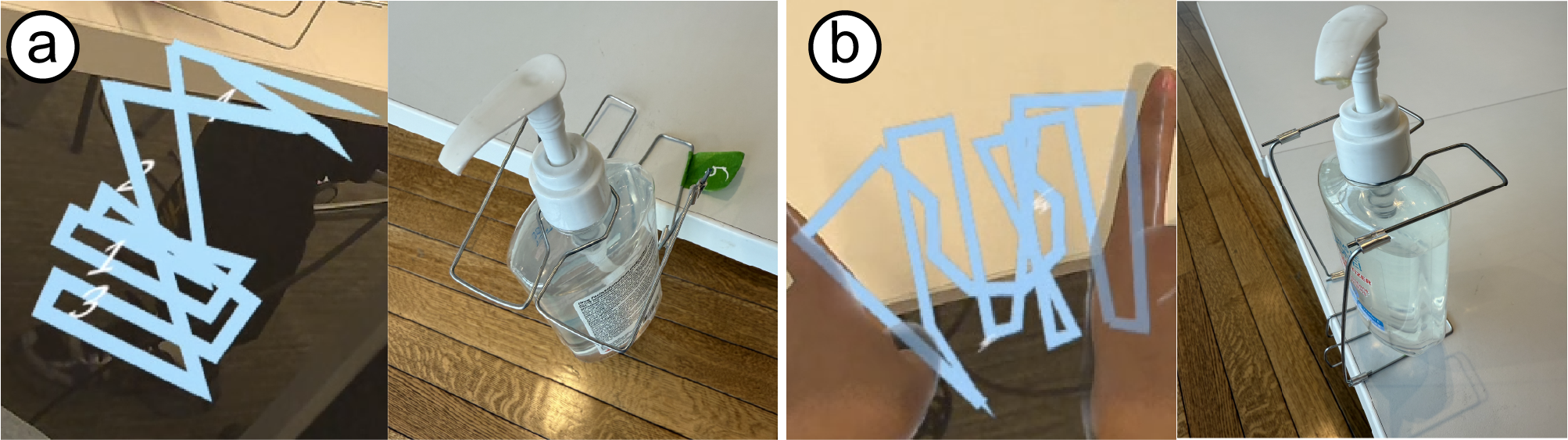}
\caption{Examples of varied arrangements in Task 1: (a) P5’s Z-shaped layered connector; (b) P10’s inward-rotated Bottle Connector.}
\label{fig:task1_variations}
\end{figure}

\paragraph{Variation in Connector Arrangements} In Task 1, participants explored different ways of arranging connectors and demonstrated notable variation in how they positioned and combined them. P5 stacked the Bottle Connector above the Table Edge Connector, forming a Z-shaped layered structure (as shown in the Fig.~\ref{fig:task1_variations} a), while P10 rotated the Bottle Connector so its opening faced inward (as shown in the Fig.~\ref{fig:task1_variations} b). These variations highlight creative reinterpretations of constrained parts, with participants varying orientation, height, and placement.

\begin{figure}[h]
\centering
\includegraphics[width=1\linewidth]{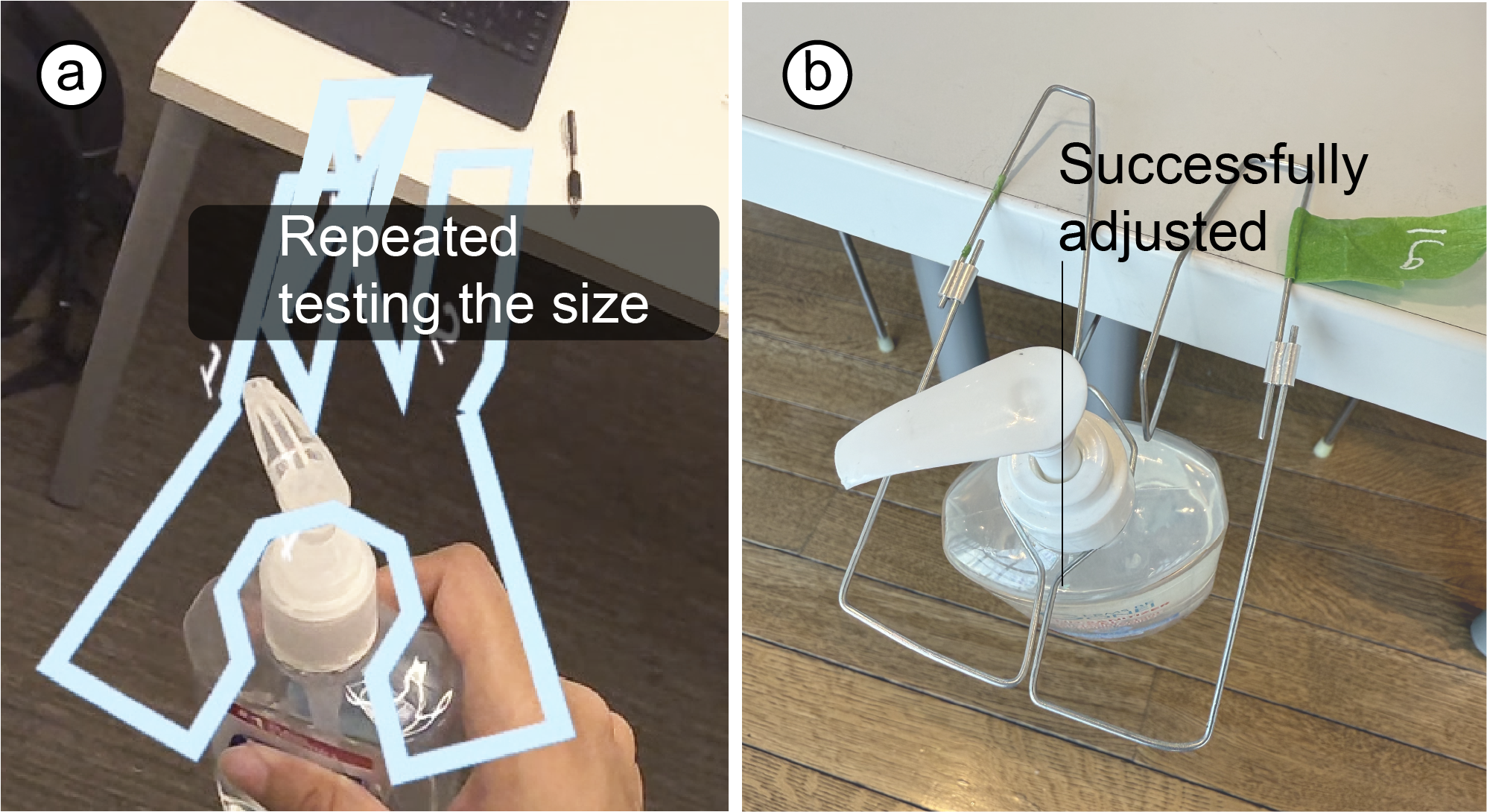}
\caption{P9’s connector: (a) repeated testing in \textit{Y-AR} before fabrication; (b) successfully adjusted fabricated result.}
\label{fig:task1_adjustment}
\end{figure}

\paragraph{Adaptability and Adjustment} Participants also leveraged the adaptability of wire structures. For example, P9 frequently compared their connector to the bottle (as shown in Fig.~\ref{fig:task1_adjustment} a) but still produced an oversized design, likely reflecting differences in spatial perception. Nevertheless, P9 successfully adjusted the connector post-fabrication to securely hold the bottle(as shown in Fig.~\ref {fig:task1_adjustment} b), illustrating the forgiving nature of wire bending and its tolerance for dimensional inaccuracies (Fig.~\ref{fig:task1_adjustment}).

\begin{figure}[h]
\centering
\includegraphics[width=1\linewidth]{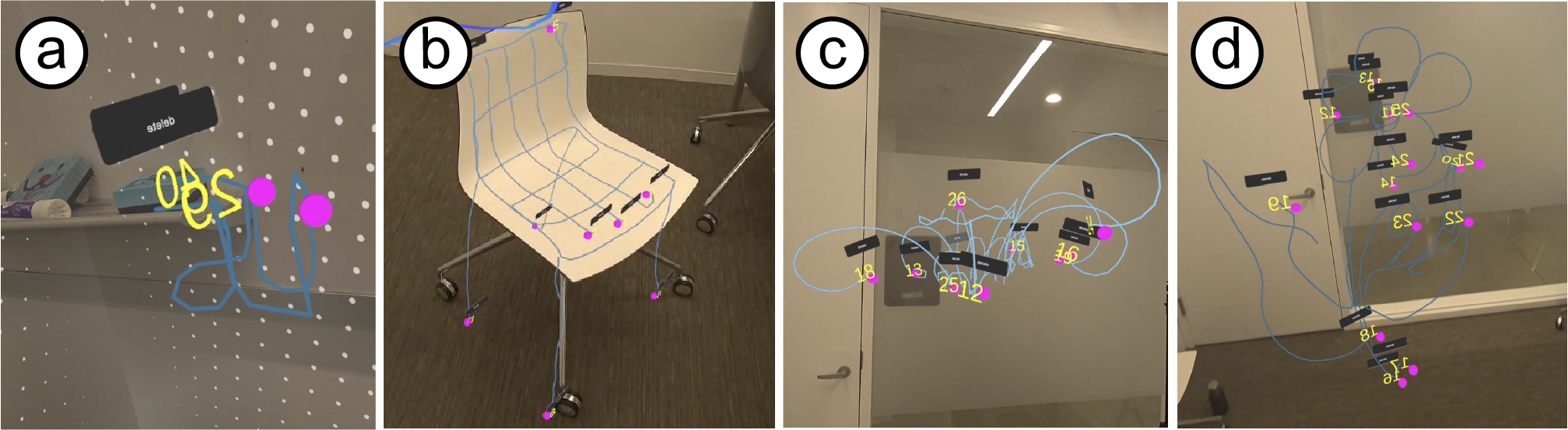}
\caption{Examples of Task 2 designs: left—environment-aware designs ((a)~P1’s hook on blackboard tray and (b)~P8’s structure on chair surface); right—expressive designs ((c)~P12’s wire crown and (d)~P6’s flower-like structure).}
\label{fig:task2_expressive}
\end{figure}

\paragraph{Environment-Aware and Expressive Design Strategies} In Task 2, participants displayed greater creative freedom, employing distinct strategies that responded to both environmental context and personal expression. Several designs explicitly incorporated surrounding features: P1 created a hook attached to the blackboard tray ledge (Fig.~\ref{fig:task2_expressive} a), while P8 constructed a structure integrated with a chair surface (Fig.~\ref{fig:task2_expressive} b). These examples illustrate how participants adapted their designs to real-world geometry, leveraging the XR environment’s spatial context. Interestingly, participants exhibited different spatial engagement strategies: some grounded their designs in the physical environment, using surrounding surfaces and contexts to guide their work, while others approached the space more abstractly, positioning forms freely without reference to environmental constraints. 

Other participants pursued expressive or artistic goals. P12, an architect, crafted an elegant wire crown using careful symmetry and looping (Fig.~\ref{fig:task2_expressive} c), while P6 designed a flower-like structure—an ambitious concept that would require additional dedicated fixtures to fabricate (Fig.~\ref{fig:task2_expressive} d).

Participants like P1 and P10 transferred structural ideas from Task 1, applying learned connector patterns to their freehand designs. Some struggled with precision due to gesture resolution limitations, which led to larger-than-intended structures.  Together, these strategies reveal \textit{Y-AR}’s potential to support both environment-aware adaptation and creative exploration, while highlighting opportunities to improve fine-detail fabrication support.

\subsection{Qualitative Findings}
We found four key topics in participants reflection on their experience. The interaction itself, the spatial context brought by XR, the design freedom, and technical limitations.

\paragraph{Intuitive interaction.}
Participants described \textit{Y-AR} as intuitive and easy to learn, with smooth gesture-function mappings that reduced the learning barrier. We computed a learnability subscale score following \citet{Lewis_Sauro_2009}, who proposed interpreting SUS items 4 and 10 as reflecting learnability. The mean learnability score was 76.0, suggesting participants generally found the system easy to learn.

P4 characterized the core interaction pattern as "pretty natural" and noted that \textit{"the logic made a lot of sense."} P10 similarly commented that although it usually takes time to understand how to use a new tool, \textit{Y-AR} was \textit{"intuitive."} P2 emphasized that gesture interactions felt smooth and well-mapped: \textit{"Each gesture has each function, and I can use the bottle to connect or use pen to draw—it’s pretty easy."}

\paragraph{Spatial context and dimensioning.}
Several participants highlighted how \textit{Y-AR} made it easier to design with real-world scale and dimensions in mind. P3 appreciated how the tool automatically aligned and measured distances: \textit{"I can hold on to any object and it kind of maps the distance, which I don’t have to measure manually—that is good."} P10 noted that \textit{"if I'm doing these gestures, it's very exact... like the dimensions."} In contrast to traditional CAD software, P11 described the difficulty of accurately imagining and measuring objects: \textit{"It’s hard to measure this one and build it precisely, and hard to imagine the real object in [red. traditional 2D CAD] modeling software."}

\paragraph{Expressive freehand design supporting creativity.}
Participants valued \textit{Y-AR}’s ability to support expressive, manual sketching without the constraints of traditional CAD interfaces. P12 described this freedom as a key benefit: \textit{"You can draw anything manually, and you can do what you want to express… you don’t have to be limited by the mouse or by software tools like the...the rules and the circle."} P2 similarly noted that compared to software like Rhino, \textit{Y-AR} was more convenient for quickly translating ideas into designs: \textit{"I can just think what I want to draw… so it’s very convenient."} Some participants found this approach inspirational; for instance, P7 did not have a clear idea initially but was influenced by the physical environment to design a flower vase: \textit{"Once I had the headset on… I just saw the table was empty… that’s why I built a flower vase with different flowers."}

\paragraph{Limitations.} Despite the overall intuitiveness of \textit{Y-AR}’s spatial interactions, some participants experienced challenges during initial use, particularly around hand-tracking precision and manipulation accuracy. P4, unfamiliar with XR, described a \textit{"learning curve"} and attributed it partly to  tracking precision. P5 similarly reported early difficulty with grabbing specific parts of objects but noted that it improved with practice: \textit{"for a specific end, it's difficult...in the beginning... But then I got used to it."} P6 also described initial interactions as \textit{"shaky," }becoming \textit{"instinctive and natural"} over time. These comments highlight that while \textit{Y-AR} lowers barriers for gesture-based design, new users still require adjustment time to achieve fine control.


\subsection{Applications.} 
We specifically primed participants by asking how and when they would be inclined to use \textit{Y-AR} in their daily lives. Two key application areas participants surfaced were augmenting existing objects, and creative ideation.

\paragraph{Customization.}
Participants widely envisioned \textit{Y-AR} supporting highly personalized designs tailored to individual needs. P4 described frustrations with off-the-shelf products like phone stands and emphasized the value of being able to \textit{"build something and choose the angle myself."} P6 noted that engineers often require specific components that are unavailable or involve long delivery times, suggesting that \textit{Y-AR} could help create \textit{"very specific pieces"} quickly in maker lab contexts. P8 proposed scenarios such as deployable camping structures or personalized jewelry, where bespoke designs are desirable.

\paragraph{Support ideation.}
Several participants identified \textit{Y-AR}’s potential at the early stages of design work, where quick iteration and open-ended form exploration are essential. Drawing on architectural experience, P3 noted its utility for tasks like steel reinforcement prototyping, describing it as well-suited to \textit{"the design iteration and ideation stage."} P7 and P9 similarly saw value for designers who need to quickly create "first drafts" or rapidly sketch and visualize ideas before refining them further in traditional CAD tools.

\subsection{Feature Requests} 

Participants provided specific feedback highlighting areas for \textit{Y-AR} improvement. 



\paragraph{Precision and control.}
Many participants requested tools to increase the precision of drawing and manipulation. They suggested features such as straight-line drawing similar to \textit{"Microsoft Paint"} (P4), linking and constraining lines (P9), and basic shape primitives like circles and rectangles for easy placement (P12). Additional requests included undo functionality (P11), improved object selection in cluttered scenes (P12), and background image tracing to guide accurate sketches (P2). Collectively, these suggestions highlight participants’ desire to complement freeform interaction with accuracy-enhancing mechanisms.


\paragraph{Bridging freeform and modular components.}
Several participants expressed interest in seamlessly combining expressive sketches with pre-designed parts, envisioning workflows that mix flexibility and structure. P1 proposed \textit{"connectors that are able to fit your custom drawings,"} while others sought CAD-like editing operations such as copy, paste, and delete (P6), and grouping behaviors to keep connected elements together (P7, P12). These ideas reflect participants’ aspirations for workflows that would enable smooth integration between freehand creativity and precise modular assembly.


\paragraph{Expanding ready-made libraries.}
Participants noted the usefulness of pre-designed components and requested a broader library of ready-made parts to speed up design. Their suggestions ranged from phone stands and laptop holders to keychains and pin holders (P2, P7, P10, P11, P12). Such libraries would provide convenient starting points for quickly building functional prototypes without requiring all geometry to be drawn from scratch.

\paragraph{Imagining future interaction possibilities.}
Some participants offered forward-looking suggestions that would extend \textit{Y-AR}’s interaction model. P3 envisioned synchronous multi-user collaboration, describing it as \textit{"like Sketchbook, but in a three-dimensional space."} P6 suggested voice commands for generating common primitives, while P12 imagined AI-assisted guidance to help users select appropriate shapes and curves based on their intent. These ideas point to opportunities for expanding \textit{Y-AR} beyond single-user, gesture-based workflows.

%% file: src/06_DISCUSSION.tex
\section{Discussion}

In this section, we reflect on key insights from the \textit{Y-AR} project, highlighting its broader implications for XR-CAD tools, wire bending, and fabrication-aware design.

\subsection{Fabrication-Aware Design CAD tool for XR}

Current CAD/CAM tools for wire bending typically provide low-level machine controls or conversion-based workflows that do not fully leverage the unique material properties and fabrication characteristics of wire structures. These tools often require technical expertise and offer limited design-focused interaction.

Our work explores how fabrication-aware design principles can be integrated into XR-CAD tools for wire bending. Rather than viewing XR tracking inaccuracy as a fundamental flaw, \textit{Y-AR} demonstrates how fabrication-aware strategies can accommodate these limitations. Specifically, \textit{Y-AR} leverages material properties such as the springiness of wire to compensate for imprecise input, allowing rough, approximate designs to still result in manufacturable, functional artifacts.

This suggests that XR CAD tools need not replicate the precision-oriented paradigms of traditional CAD systems; instead, they might offer value by supporting design approaches that are inherently robust to variation and uncertainty. Importantly, this approach is extensible beyond wire bending: many CAD-to-fabrication workflows face uncertainties that lead to discrepancies between design and physical outcome. Adaptive structures—those that tolerate or even leverage imprecision—can help bridge this gap by turning fabrication uncertainty into certainty. Future XR-CAD systems could embed tolerance-aware design suggestions or guide users toward structures that naturally accommodate fabrication imprecision, making such adaptability a core design strategy.

\subsection{Gesture Discoverability}

Another contribution of \textit{Y-AR} is its support for gesture discoverability. Users did not need to memorize abstract symbolic gestures but could enact natural, mimetic gestures that closely mirror real-world actions (e.g., tracing around a table edge or pointing to a bottle rim).

We observed that participants with little prior XR experience were nonetheless able to quickly adapt to \textit{Y-AR}’s gesture-based interface. This suggests that gesture interaction provides a natural and accessible way for users to express object relationships in XR environments, similar to metaphors in games (e.g., the intuitive slicing gesture in Fruit Ninja). Our findings indicate that mimetic gesture design allows effective operation without extensive training or memorization.

\subsection{Gestural Adaptation in Connector Design}

Prior connector design tools, such as \textit{AutoConnect}~\cite{10.1145/2816795.2818060}, primarily rely on predefined geometric shapes (e.g., cylinders, boxes) to generate connection solutions. While this approach appears general, real-world objects exhibit far greater complexity and diversity than these primitives. Most everyday objects feature irregular contours, compound surfaces, or non-standard geometries, limiting the applicability of methods based on standard shapes. Under this shape-to-connector mapping paradigm, the design space becomes constrained by geometric limitations.

\textit{Y-AR} transforms this landscape through gesture-based interaction. Human hands can naturally trace and conform to complex object contours, from curved bottles to composite surfaces. This shifts the process from "shape matching" to "functional exploration." Moreover, gesture interaction no longer produces singular, deterministic connection solutions. Instead, users can explore multiple connection possibilities for the same object via different gestural paths, gripping strategies, and tracing approaches. For example, users can employ overhead suspension gestures for load-bearing connectors, bottom-up supporting gestures for structural supports, or lateral fixation gestures for clamping mechanisms.

This transition—from constraint to exploration, determinism to diversity—dramatically enriches the design space. Unlike tools enforcing rigid geometric correspondence, \textit{Y-AR}'s gesture-based approach enables a one-to-many relationship between objects and solutions, where the same physical object yields fundamentally different connector designs depending on the intended interaction mode—each discovered through natural, embodied exploration rather than predetermined geometric analysis.

\subsection{Future Work}

We found that wire bending is rarely used in isolation but often serves as one component in multi-material fabrication. Wire structures frequently act as frames or supports, combined with other techniques such as 3D printing for joints, laser cutting for panels, or fabric wrapping for surfaces to achieve complete, functional artifacts.

This highlights an opportunity for future CAD tools to support multi-workflow integration, providing an interactive environment where users can holistically plan, visualize, and refine hybrid fabrication workflows. By supporting combinations of additive, subtractive, and formative manufacturing processes, such tools could improve the functionality of their designs.

At the same time, we note that wire bending itself remains underexplored in personal fabrication. Current CAD/CAM tools provide little support for wire-based structures, limiting novices' ability to explore this design space.

\textit{Y-AR} represents a first step toward addressing this gap. Future research could expand the wire structure design space by enabling more complex geometries, for example through generative design features that provide sophisticated starting points or automated path optimization tools that reduce manual trial-and-error when planning manufacturable geometries.

\section{CONCLUSION}

We presented \textit{Y-AR}, a fabrication-aware design CAD tool for 3D wire bending. This fabrication-aware environment allows designers to create well-designed wire-bent structures. We consider our guidelines for designing for wire bending to be valuable as contributions, independent of the specific implementation of the \textit{Y-AR} environment. A central insight of our approach is that the springiness of the material can be used to overcome modeling inaccuracies. This allows \textit{Y-AR} to rely on much less precise input techniques such as mid-air gestures in XR. The XR environment, in turn, supports more context-aware design by helping users create objects that fit their physical surroundings. We believe this core idea will allow developers to build fabrication-aware CAD tools for fabrication techniques that extend beyond wire bending. Our user evaluation indicates that \textit{Y-AR} allows users to create functional structures with wire bending that would have been hard to accomplish with existing CAM tools. 

\textit{Y-AR} will be available as an open-source tool for other researchers to extend. Future work could explore the integration of novel tool types or the expansion of design capabilities to accommodate fabrication methods beyond wire bending. Related work has shown that wire bending is well equipped to provide structure for other fabrication processes. We are also interested in running more longer-term deployment to see how users benefit from \textit{Y-AR} over an extended amount of time. And as stated before, the framework of \textit{Y-AR}, combining XR and springiness, offers a promising foundation for other future fabrication-aware design environments.

%% file: src/08_appendix.tex
\appendix

\section{Participant Demographics}

\begin{table*}[h]
\centering
\caption{Overview of study participants.}
\begin{tabular}{p{0.4cm}p{1.8cm} p{3.2cm} p{3cm} p{3.2cm}}\toprule
\textbf{ID} & \textbf{Age Group} & \textbf{XR usage} & \textbf{CAD usage} & \textbf{Background} \\ 
\midrule
P1 & Under 18 & Intermediate & Regularly & FIRST robotics Team\\
P2 & 25–34 & Occasionally & Regularly & Engineer student \\
P3 & 25–34 & Rarely & Regularly & Architecture \\
P4 & 18–24 & Rarely & Never & CS student \\
P5 & 18–24 & Once or twice & Occasionally & CS student \\
P6 & 18–24 & Rarely & Occasionally & CS student \\
P7 & 18–24 & Never & Once or twice & CS student \\
P8 & 25–34 & Rarely & Regularly & Design student \\
P9 & 18–24 & Rarely & Occasionally & Engineer student \\
P10 & 18–24 & Rarely & Once or twice & CS student \\
P11 & 18–24 & Rarely & Never & Design student\\
P12 & 18–24 & Rarely & Regularly & Architecture\\ 
\bottomrule
\end{tabular}
\label{tab:expert-participants}
\end{table*}